\documentclass[prd,aps,floats,epsfig,twocolumn]{revtex4}  

\usepackage{enumitem}
\usepackage{graphicx}
\usepackage{subfigure}
\usepackage{amssymb}
\usepackage{amsmath}
\usepackage{bbm}
\usepackage{color}

\def\bea{\begin{eqnarray}}
\def\eea{\end{eqnarray}}
\begin{document}

\input{epsf}

\title{Cosmological constraints on general, single field inflation}

\author{Nishant Agarwal}
\author {Rachel Bean}

\affiliation{Department of Astronomy, Cornell University, Ithaca, NY 14853, USA.}

\begin{abstract}
{Inflation is now an accepted paradigm in standard cosmology, with its predictions consistent with observations of the cosmic microwave background. It lacks, however, a firm physical theory, with many possible theoretical origins beyond the simplest, canonical, slow-roll inflation, including Dirac-Born-Infeld inflation and k-inflation. We discuss how a hierarchy of Hubble flow parameters, extended to include the evolution of the inflationary sound speed, can be applied to compare a general, single field inflationary action with cosmological observational data. We show that it is important to calculate the precise scalar and tensor primordial power spectra by integrating the full flow and perturbation equations, since values of observables can deviate appreciably from those obtained using typical second-order Taylor expanded approximations in flow parameters. As part of this, we find that a commonly applied approximation for the tensor to scalar ratio, $r \approx 16 c_s\epsilon$, becomes poor (deviating by as much as 50\%) as $c_s$ deviates from 1 and hence the Taylor expansion including next-to-leading order contribution terms involving $c_{s}$ is required. By integrating the full flow equations, we use a Monte-Carlo-Markov-Chain approach to impose constraints on the parameter space of general single field inflation, and reconstruct the properties of such an underlying theory in light of recent cosmic microwave background and large-scale structure observations.} 

\end{abstract}
\maketitle

%%%%%%%%%%%%%%%%%%%%%%%%%%%%%%%%%%%%%%%%%%%%%%
\section{Introduction}
\label{One}
%%%%%%%%%%%%%%%%%%%%%%%%%%%%%%%%%%%%%%%%%%%%%%

Even though inflation explains our observable universe remarkably, we have very little understanding of the physical mechanism responsible for the acceleration during inflation. Finding a physical basis for inflation is likely to help in understanding particle physics at very high energies. Many models of inflation are motivated by supergravity, the string landscape and D-branes \cite{Dvali:1998pa,Wehus:2002se,Kachru:2003sx,Cline:2006hu,HenryTye:2006uv}. Therefore understanding inflation may also be useful in testing string theory \cite{Shandera:2006ax,Easson:2007dh,Bean:2007eh,Bean:2007hc,Peiris:2007gz}. 

Recent advances in precision cosmology provide valuable constraints on the cosmological density perturbation, which is essential to understand the inflationary scenario. Improved measurements of the temperature and polarization anisotropies of the cosmic microwave background (CMB) \cite{Komatsu:2008hk,Dunkley:2008ie,Nolta:2008ih,Gold:2008kp,Hill:2008hx,Hinshaw:2008kr} and data from large scale structure surveys \cite{Tegmark:2003ud,Tegmark:2003uf,Cole:2005sx,Sanchez:2005pi,Tegmark:2006az} together characterize the primordial spectrum of fluctuations to fine detail.

Using observations to constrain the primordial power spectrum one can reconstruct properties of the underlying theory guiding the physics of the inflationary era \cite{Salopek:1989,Copeland:1993ie,Copeland:1993zn,Adams:1994qp,Turner:1995ge,Grivell:1999wc,Hoffman:2000ue,Adams:2001vc,Easther:2002rw,Kinney:2003uw,Chung:2003iu,Makarov:2005uh,Kadota:2005hv,Chung:2005hn,Ballesteros:2005eg,Peiris:2006ug,Cline:2006db,Cortes:2006ap,Martin:2006rs,Ballesteros:2007te,Lorenz:2007ze,Boyanovsky:2007ry,Powell:2007gu,Lesgourgues:2007aa,Bean:2008ga,Adshead:2008vn,Gauthier:2008mq,Lorenz:2008je}. In order to consider what observations tell us, without any theoretical bias, we need to reconstruct the entire inflaton action, instead of just the inflaton potential or a specific kinetic term, since theories of inflation, such as those arising from the Dirac-Born-Infeld (DBI) action \cite{Dvali:1998pa} or from k-inflation \cite{ArmendarizPicon:1999rj,Garriga:1999vw,ArmendarizPicon:2000dh,ArmendarizPicon:2000ah}, allow the presence of nonminimal kinetic terms. A hierarchy of derivatives of the Hubble expansion factor, ``flow parameters", during inflation was developed as a technique to reconstruct canonical inflation \cite{Copeland:1993zn,Liddle:1994dx,Kinney:2002qn,Liddle:2003py}. This was recently extended to DBI inflation by also considering derivatives of the inflaton sound speed  \cite{Peiris:2007gz} and, through an additional derivative of the Lagrangian, to a general single field action in \cite{Bean:2008ga}.

The paper is organized as follows. In Sec. \ref{Two} we review the background evolution equations and the flow formalism in general, single field inflation. In Sec. \ref{Three} we discuss how we calculate the exact primordial scalar and tensor perturbation spectra by integrating the flow equations, and review the ability of approximate Taylor expansions about a pivot point to describe physical observables such as the tilt, running and tensor to scalar ratio in the general inflationary scenario. In Sec. \ref{Four} we present the main findings of the paper, cosmological constraints on the general, single field inflationary action in light of current CMB temperature and polarization power spectra and three point temperature correlation, large-scale structure power spectrum and supernovae luminosity distance constraints. We consider constraints within the observed range of physical scales $10^{-4}$ Mpc$^{-1}$ $\lesssim k\lesssim 1$ Mpc$^{-1}$ in Sec. \ref{Fourb}, as well as general action reconstruction over the extended inflationary history in Sec. \ref{Fourc}. In Sec. \ref{Five} we draw together our findings and discuss implications for the future.

%%%%%%%%%%%%%%%%%%%%%%%%%%%%%%%%%%%%%%%%%%%%%%
\section{The Hubble flow formalism}
\label{Two}
%%%%%%%%%%%%%%%%%%%%%%%%%%%%%%%%%%%%%%%%%%%%%%

Consider the general lagrangian $\mathcal{L}(X,\phi)$ of a single scalar field inflationary model. Here $X = \frac{1}{2}\partial_{\mu}\phi\partial^{\mu}\phi$ is the canonical kinetic term. The pressure and energy density are given by,
\begin{eqnarray}
	p(X,\phi) & \equiv & \mathcal{L}(X,\phi), \\
	\rho(X,\phi) & \equiv & 2X\mathcal{L}_{X} - \mathcal{L}(X,\phi),
\end{eqnarray}
where $\mathcal{L}_{X} \equiv \partial\mathcal{L} / \partial X$. We assume that the null energy condition $\rho+p>0$, is satisfied, such that,
\begin{eqnarray}
	\mathcal{L}_{X} > 0.
\end{eqnarray}

The adiabatic sound speed for the propagation of inhomogeneities, $c_{s}$, is defined as,
\begin{equation}
	c_{s}^{2} \equiv \frac{p_{X}}{\rho_{X}} = \left( 1 + 2\frac{X\mathcal{L}_{XX}}{\mathcal{L}_{X}} \right)^{-1}.
\end{equation}
We measure the extent of inflation using the variable $N_{e}$, which denotes the number of e-folds before the end of inflation. We choose $N_{e}$ to increase backwards in time from the end of inflation, i.e.,
\begin{eqnarray}
	dN_{e} & = & -Hdt, \\
	N_{e}  & \equiv & \ln \frac{a(t_{en})}{a(t)},
\end{eqnarray}
where $a(t)$ is the scale factor at any time $t$, and $t_{en}$ is the time at the end of inflation. 

We can define three physical slow-roll parameters to describe time derivatives of the Hubble parameter and sound speed,
\bea
	\epsilon &\equiv&  -\frac{\dot{H}}{H^2}, \hspace{0.5cm} \eta \equiv \frac{\dot{\epsilon}}{H\epsilon} , \hspace{0.5cm} \kappa \equiv -\frac{\dot{(c_s^{-1})}}{H c_s^{-1}}, 
\eea
where a dot represents a derivative with respect to time, $t$. Note that these parameters are independent of a scalar field definition. They depend upon $\mathcal{L}$, and combinations of $X$ and derivatives of $\mathcal{L}$ with respect to $X$ and $\phi$ that are invariant under a scalar field redefinition. The acceleration equation can now be written as
\begin{equation}
	\frac{\ddot{a}}{a} = (1-\epsilon) H^{2}, \label{acceleq}
\end{equation}
requiring $\epsilon \le 1$ for inflation to occur. 

The slow-roll approximation requires that
\begin{eqnarray}
	\epsilon, \eta, \kappa, \epsilon_{N}, \eta_{N}, \kappa_{N}, ... \ll 1,	
\end{eqnarray}
where $\epsilon_{N} \equiv d\epsilon/dN_{e}$, etc.

In order to describe an action beyond the slow-roll assumption, one can define an infinite hierarchy of ``flow parameters", as used extensively for canonical inflation \cite{Liddle:1994dx,Kinney:2002qn,Liddle:2003py} and extended to DBI inflation \cite{Peiris:2007gz}, and to a general action in \cite{Bean:2008ga}. For a general action, with a general scalar field definition, the evolution is described by three hierarchies of the flow parameters, dealing with derivatives with respect to the scalar field of the Hubble constant ($H$), the speed of sound ($c_{s}$), and $\mathcal{L}_{X}$. These parameters are in general all dependent on the explicit choice of $\phi$  and as discussed in \cite{Bean:2008ga}, actions reconstructed using this formalism can map onto each other through a scalar field redefinition.  In this paper we impose a specific scalar field choice, such that $\mathcal{L}_{X}=c_{s}^{-1}$, consistent with canonical and DBI inflation, to alleviate this degeneracy. This leaves us with only two distinct hierarchies of flow parameters,
\begin{eqnarray}
	\epsilon & = & \frac{2M_{pl}^{2}}{c_{s}^{-1}} \left( \frac{H'}{H} \right)^{2}, \\
	\kappa & = & \frac{2M_{pl}^{2}}{c_{s}^{-1}} \left( \frac{H'}{H} \frac{(c_{s}^{-1})'}{c_{s}^{-1}} \right),
\end{eqnarray}
and
\begin{eqnarray}
	^{l}\lambda(\phi) & = & \left( \frac{2M_{pl}^{2}}{c_{s}^{-1}} \right)^{l} \left( \frac{H'}{H} \right)^{l-1} \frac{H^{[l+1]}}{H}, \\
	^{l}\alpha(\phi) & = & \left( \frac{2M_{pl}^{2}}{c_{s}^{-1}} \right)^{l} \left( \frac{H'}{H} \right)^{l-1} \frac{(c_{s}^{-1})^{[l+1]}}{c_{s}^{-1}}, 
\end{eqnarray}
for $l \geq 1$. Here a prime denotes derivative with respect to $\phi$, $M_{pl}^{2} = 1/8\pi G$, and $H^{[l+1]} \equiv d^{l+1}H/d\phi^{l+1}$ etc. The combination of parameters, $2{^1\lambda}-\kappa = 2\epsilon-\eta$, is invariant under scalar field redefinition.

Using
\begin{equation}
	\frac{d\phi}{dN_{e}} = \frac{2M_{pl}^{2}}{\mathcal{L}_{X}} \frac{H'}{H},
\end{equation}
we can write the evolutionary paths of the flow parameters as a set of coupled first order differential equations with respect to $N_{e}$, 
\begin{eqnarray}
	\epsilon_{N} & = & -\epsilon (2\epsilon - 2{^1\lambda} + \kappa) = -\epsilon \eta, \label{flowstart} \\
	\kappa_{N} & = & -\kappa(\epsilon - {^1\lambda} + 2\kappa) + \epsilon {^1\alpha},
\end{eqnarray}
and for $l \geq 1$,
\begin{eqnarray}
	^{l}\lambda_{N} & = & -^{l}\lambda [l\epsilon - (l-1){^1\lambda} + l\kappa] + {^{l+1}\lambda}, \\
	^{l}\alpha_{N} & = & -^{l}\alpha [(l-1)\epsilon - (l-1){^1\lambda} + (l+1)\kappa] + {^{l+1}\alpha}. \ \ \ \ \ \  \label{flowend} 
\end{eqnarray}

In this paper we consider two scenarios in which inflation is driven by the inflationary flow equations, one in which the end of inflation arises from  when $\epsilon=1$, and one in which inflation does not end on its own ($\epsilon\neq 1$), but may be brought on, for example, by the behavior of a second scalar field. 

%%%%%%%%%%%%%%%%%%%%%%%%%%%%%%%%%%%%%%%%%%%%%%
\section{Primordial perturbations}
\label{Three}
%%%%%%%%%%%%%%%%%%%%%%%%%%%%%%%%%%%%%%%%%%%%%%

In this section we discuss the generation of primordial power spectra in single field inflation: we summarize the evolution equations for the scalar and tensor perturbations in Sec. \ref{Threea}, the choice of initial conditions in Sec. \ref{Threeb}, and how the exact power spectra are calculated through evolving the flow equations in Sec. \ref{Threec}. We also review the approximate expressions for the power spectra in terms of the flow parameters in order to compare them with the exact power spectra we use for the analysis in Sec. \ref{Four}.

%%%%%%%%%%%%%%%%%%%%%%%%%%%%
\subsection{Calculating the power spectrum}
\label{Threea}
%%%%%%%%%%%%%%%%%%%%%%%%%%%%

The evolution of the scalar perturbations in the metric,
\begin{eqnarray}
	ds^{2} = (1+2\Phi)dt^{2} - (1-2\Phi)a^{2}(t)\gamma_{ij}dx^{i}dx^{j},
\end{eqnarray}
are typically described in terms of the Bardeen parameter, $\zeta$, 
\bea 
	\zeta = \frac{5\rho+3p}{3(\rho+p)}\Phi + \frac{2\rho}{3(\rho+p)}\frac{\dot{\Phi}}{H},
\eea
and specifically its spectral density,
\bea
	\mathcal{P}_{\zeta} & = & \frac{k^{3}}{2\pi^{2}} \zeta^2,
 	\label{scalarps}
\eea
while the tensor perturbations can be characterized by a metric with $g_{00}=-1$, zero space-time components $g_{0i}=0$, and $\delta g_{ij}=h_{ij}$. We can decompose these perturbations into two independent polarization modes, denoted $+$ and $\times$, since gravitational waves are both transverse and traceless. Writing the Fourier modes as $h_{k,+}$ and $h_{k,\times}$, the spectral density of tensor fluctuations, $\mathcal{P}_{h}$ can be written as,
\begin{eqnarray}
	\mathcal{P}_{h} = \frac{k^{3}}{2\pi^{2}} \left( \left\langle |h_{k,+}|^{2} \right\rangle + \left\langle |h_{k,\times}|^{2} \right\rangle \right).
\end{eqnarray}
The evolution of $\zeta$ and $h_{\pm}$ can be calculated concisely through considering two alternative Mukhanov variables, 
\bea
	u_{k} &\equiv & z \zeta,
 	\label{uk} \\
 	v_{+,\times} &\equiv& \left( \frac{aM_{pl}}{2} \right) h_{+,\times},
\eea
where
\begin{eqnarray}
	z & = & \frac{a(\rho+p)^{1/2}}{c_{s}H} = \frac{\sqrt{2}M_{pl}a\sqrt{\epsilon}}{c_{s}}.
	\label{zeqn}
\end{eqnarray}

To determine the full evolution of the power spectrum, we need to numerically integrate the mode equations in $u_{k}$ and $v_{k}$. Written in terms of the number of e-foldings $N_e$, these are
\begin{eqnarray}
	\frac{d^{2}u_{k}}{dN_{e}^{2}} - (1-\epsilon)\frac{du_{k}}{dN_{e}} + \left[ \left(\frac{c_{s}k}{aH}\right)^{2} - W \right] u_{k} = 0,
\label{scalarmodeqn2} \\
	\frac{d^{2}v_{k}}{dN_{e}^{2}} - (1-\epsilon)\frac{dv_{k}}{dN_{e}} + \left[ \left(\frac{k}{aH}\right)^{2} - (2-\epsilon) \right] v_{k} = 0,
\label{tensormodeqn2}
\end{eqnarray}
with, 
\bea
	W & = & 2\left[ \left( 1+\frac{\eta}{2}-\kappa \right) \left( 1-\frac{\epsilon}{2}+\frac{\eta}{4} - \frac{\kappa}{2} \right) \right] \nonumber \\
	& & + \frac{\eta_{N}}{2} - \kappa_{N}.  \ \ \ \ \ \
\eea

Following \cite{Chen:2006nt,Kinney:2007ag}, the scalar spectral density, $\mathcal{P}_{\zeta}$ is given by
\begin{eqnarray} 
	\mathcal{P}_{\zeta} = 2^{2\nu-3} \left| \frac{\Gamma(\nu)}{\Gamma(3/2)} \right|^{2} (1-\epsilon-\kappa)^{2\nu-1} \left| \frac{H^{2}}{2\pi\sqrt{2X}} \right|^{2}_{c_{s}k = aH}, \nonumber \\ 
	\label{scalps1}
\end{eqnarray}
where
\begin{eqnarray}
	\nu & = & \frac{3}{2} + \epsilon + \frac{\eta}{2} + \frac{\kappa}{2}.
\end{eqnarray} 
The tensor spectral density is
\begin{eqnarray} 
	\mathcal{P}_{h} = 2^{2\mu-3} \left| \frac{\Gamma(\mu)}{\Gamma(3/2)} \right|^{2} (1-\epsilon)^{2\mu-1} \left| \frac{\sqrt{2}H}{\pi M_{pl}} \right|^{2}_{k = aH}, \nonumber \\
	\label{tensorps1}
\end{eqnarray}
where
\begin{eqnarray}
	\mu^{2} & = & \frac{2-\epsilon}{(1-\epsilon)^{2}} + \frac{1}{4}.
\end{eqnarray}
	  
%%%%%%%%%%%%%%%%%%%%%%%%%%%%
\subsection{Scalar and tensor perturbation initial conditions}
\label{Threeb}
%%%%%%%%%%%%%%%%%%%%%%%%%%%%

We assume the standard choice of initial conditions for the mode functions $u_{k}$ and $v_{k}$, the Bunch-Davies vacuum,
\begin{eqnarray}	
	& & u_{k}(-c_{s}k\tau \rightarrow \infty) = \frac{1}{\sqrt{2c_{s}k}} e^{-ic_{s}k\tau},
	\label{scalarbunchdavies} \\
	& & v_{k}(-k\tau \rightarrow \infty) = \frac{1}{\sqrt{2k}} e^{-ik\tau}.
	\label{tensorbunchdavies}
\end{eqnarray}
As in \cite{Powell:2007gu} we note that we cannot use these conditions directly in order to solve the mode equations numerically since we cannot impose these conditions in the infinite past. We need to initialize the mode functions at sufficiently early times, which we choose as the number of e-folds before the end of inflation at which $(c_{s}k/aH)/(1-\epsilon-\kappa) = 50$ for scalar perturbations, and $(k/aH)/(1-\epsilon) = 50$ for tensor perturbations. Note that the results are insensitive to the precise condition chosen.  

We define the ratio of the Hubble radius to the proper wavelength of fluctuations for the scalar and tensor perturbations, respectively, as
\begin{eqnarray}	
	 y_{\zeta}(N_e) &\equiv& \frac{c_{s}k}{aH}, \\
	 y_{h}(N_e) &\equiv &\frac{k}{aH}.
\end{eqnarray}
Then,
\begin{eqnarray}
	\frac{dy_{\zeta}}{d\tau} & = & -c_{s}k(1-\epsilon-\kappa), \\
	\frac{d\epsilon}{dy_{\zeta}} & = & \frac{1}{y_{\zeta}(1-\epsilon-\kappa)} \frac{d\epsilon}{dN_{e}}, \\
	\frac{d\kappa}{dy_{\zeta}} & = & \frac{1}{y_{\zeta}(1-\epsilon-\kappa)} \frac{d\kappa}{dN_{e}},
\end{eqnarray}
and
\begin{eqnarray}
	\frac{dy_{h}}{d\tau} & = & -k(1-\epsilon), \\
	\frac{d\epsilon}{dy_{h}} & = & \frac{1}{y_{h}(1-\epsilon)} \frac{d\epsilon}{dN_{e}}.
\end{eqnarray}
The initial conditions for each mode need to be set at early times that correspond to large $y_{\zeta}$ and $y_{h}$, when the scalar/tensor mode is well within the horizon. We see from the above equations that at large $y_{\zeta}$ and $y_{h}$, $\epsilon(y_{\zeta})$, $\kappa(y_{\zeta})$, and $\epsilon(y_{h})$ are approximately constant. Then we can integrate the equations in $y_{\zeta}$ and $y_{h}$ to get,
\begin{eqnarray}
	&& y_{\zeta} = -c_{s}k\tau(1-\epsilon-\kappa), \\
	&& y_{h} = -k\tau(1-\epsilon).
\end{eqnarray} 
Using this in (\ref{scalarbunchdavies}) and (\ref{tensorbunchdavies}) we get the initial conditions,
\begin{eqnarray}
	& & u_{k}(y_{\zeta i}) = \frac{1}{\sqrt{2c_{s}k}} e^{iy_{\zeta i}/(1-\epsilon_{i}-\kappa_{i})}, \\
	& & \frac{du_{k}}{dN_{e}}\bigg|_{y_{\zeta}=y_{\zeta i}} = \frac{i}{\sqrt{2c_{s}k}} y_{\zeta i} e^{iy_{\zeta i}/(1-\epsilon_{i}-\kappa_{i})},
\end{eqnarray} 
and,
\begin{eqnarray}
	& & v_{k}(y_{hi}) = \frac{1}{\sqrt{2k}} e^{iy_{hi}/(1-\epsilon_{i})}, \\
	& & \frac{dv_{k}}{dN_{e}}\bigg|_{y_{h}=y_{hi}} = \frac{i}{\sqrt{2k}} y_{hi} e^{iy_{hi}/(1-\epsilon_{i})}.
\end{eqnarray} 

%%%%%%%%%%%%%%%%%%%%%%%%%%%%
\subsection{Calculating the primordial power spectrum}
\label{Threec}
%%%%%%%%%%%%%%%%%%%%%%%%%%%%

We set the initial conditions for each scalar $k$-mode at the number of e-folds, $N_{e}$, for which $y_{\zeta i}/(1-\epsilon_{i}-\kappa_{i}) = 50$. We then integrate the mode equation (\ref{scalarmodeqn2}) to find $u_{k}(N_{e})$ as we go forward in time. At each instant we obtain the value of $z$ from (\ref{zeqn}). The flow parameters are simultaneously integrated using their first order differential equations (\ref{flowstart})-(\ref{flowend}), as are the speed of sound and $aH$, using
\begin{eqnarray}
	\frac{dc_{s}}{dN_{e}} & = & -\kappa c_{s}, \ \ \ 
	\frac{d(aH)}{dN_{e}}  =  -(1-\epsilon)aH.
\end{eqnarray}
We find the scalar power spectrum (for each value of $k$, using $\zeta$ from equation (\ref{uk}) in (\ref{scalarps})) as we evolve forward in time until the power spectrum freezes out, the condition for which we set as $[d \ln \mathcal{P}_{\zeta}/d\ln a] < 10^{-3}$. At this level the accuracy of the power spectrum calculation is at least as good as the accuracy of the other numerical calculations required when obtaining the CMB and matter power spectra predictions in CAMB, described in Sec. \ref{Four}. This allows efficient computational calculation of the primordial spectrum at a level of accuracy sufficient not to degrade the overall accuracy of the cosmological predictions obtained using the CAMB code, as described in Sec. \ref{Four}. Similarly for each tensor $k$-mode we set the initial conditions at $N_{e}$ for which $y_{hi}/(1-\epsilon_{i}) = 50$ and find the power spectrum as we evolve forward in time until it freezes out at $[d \ln \mathcal{P}_{h}/d\ln a] < 10^{-3}$. Note therefore that we do not simply evaluate the power spectrum at horizon crossing, defined as $c_sk=aH$, as we discuss below. Assuming such an instantaneous freeze-out takes place at this time, and further assuming that both tensor and scalar modes freeze out nearly simultaneously can have notable effects on the estimation of the power spectrum  variables.

To first order in slow roll parameters, the equation for the scalar power spectrum (\ref{scalps1}) becomes \cite{Chen:2006nt,Kinney:2007ag},
\begin{eqnarray}
	\mathcal{P}_{\zeta}(k) & = & \left[ 1-2\epsilon -2\kappa+ 2b \left( \epsilon+\frac{\eta}{2}+\frac{\kappa}{2} \right) \right] \nonumber \\ 
	& & \times \ \frac{1}{8\pi^{2}M_{pl}^{2}} \frac{H^{2}}{c_{s}\epsilon} \bigg|_{c_{s}k = aH},
\label{Pzeta}
\end{eqnarray}
where $b=2-\ln 2-\gamma$, and $\gamma = 0.5772$ is the Euler-Mascheroni constant. A similar calculation for the tensor power spectrum (\ref{tensorps1}) is typically evaluated at tensor mode horizon crossing $(k=aH)$, however recently an approximate expression at scalar horizon crossing $(c_sk=aH)$ was also given \cite{Lorenz:2008je,Lorenz:2008et},
\begin{eqnarray}
	\mathcal{P}_{h}(k) & = & \left[ 1-2(1-b)\epsilon \right] \frac{2H^{2}}{\pi^{2}M_{pl}^{2}} \bigg|_{k = aH} \\
	& \approx & \left[ 1-2(1-b-\ln c_{s})\epsilon \right] \frac{2H^{2}}{\pi^{2}M_{pl}^{2}} \bigg|_{c_{s}k = aH}.
\label{Phatshc}
\end{eqnarray}
We consider pivot scales, at which spectrum parameters are calculated, for scalar and tensor modes as $k_{*s}=0.01$ Mpc$^{-1}$ and $k_{*t}=0.01$ Mpc$^{-1}$, respectively. 

Using the above expressions one can calculate the scalar power spectrum normalization, $A_s$, tilt, $n_s$, and running, $n_{run}$, of the spectral index, and the tensor spectral index, $n_{t}$, at the pivot points, \cite{Kinney:2007ag,Lorenz:2008et}
\begin{eqnarray}
	A_{s} & \equiv & \mathcal{P}_{\zeta}(k_{*s}), 
\label{normalization}
\end{eqnarray}
\begin{eqnarray}
	n_{s} - 1 & \equiv & \frac{d \ln \mathcal{P}_{\zeta}}{d \ln k} \bigg|_{k=k_{*s}} 
\label{nsexact} \\
	& \approx & - (2\epsilon + \eta + \kappa)(1+\epsilon+\kappa) \nonumber \\
	& & - 2b \left( \epsilon_{N} + \frac{\eta_{N}}{2} + \frac{ \kappa_{N}}{2} \right)+ 2\epsilon_{N} + 2\kappa_N, 
\label{nsapprox}
\end{eqnarray}
\begin{eqnarray}
	n_{run} & \equiv & \frac{d n_{s}}{d \ln k} \bigg|_{k=k_{*s}}
\label{nrunexact} \\
	& \approx & (2\epsilon + \eta + \kappa)(\epsilon_{N} + \kappa_{N}) + (2\epsilon_{N} +\eta_{N} + \kappa_{N}) \nonumber \\
	& & \times \ (1+\epsilon+\kappa)^2 + 2b \left( \epsilon_{NN} +\frac{\eta_{NN}}{2} + \frac{ \kappa_{NN}}{2} \right) \nonumber \\
	& & - 2\epsilon_{NN} - 2\kappa_{NN}, \ \ \ \ \
\label{nrunapprox}
\end{eqnarray}

\begin{eqnarray}
	n_{t} & \equiv & \frac{d \ln \mathcal{P}_{h}}{d \ln k} \bigg|_{k=k_{*t}} 
\label{ntexact} \\
	& \approx & [-2\epsilon(1+\epsilon+\kappa) + 2(1-b)\epsilon_{N}]_{k=k_{*t}} \\
	& \approx & [-2\epsilon(1+\epsilon+\kappa) + 2(1-b-\ln c_{s})\epsilon_{N} + 2\epsilon\kappa]_{k=k_{*s}}, \nonumber \\
\label{ntapprox} 
\end{eqnarray}
where the approximate expressions in terms of the slow-roll parameters are given to second order for $n_{s}$ and $n_{t}$, and third order for $n_{run}$. All parameters are calculated at sound horizon crossing, unless stated otherwise. 

The Taylor expanded expressions for the scalar and tensor power spectrum, spectral indices and running reduce to previous results for canonical inflation \cite{Kinney:2002qn,Leach:2002ar,Habib:2004kc,Peiris:2006ug,Hamann:2008pb} and for general inflation \cite{Garriga:1999vw,Kinney:2007ag,Bean:2008ga,Lorenz:2008et} to the orders quoted in those papers.  

We define the tensor-to-scalar ratio without Taylor expansion approximations as $r_{exact}$, 
\begin{eqnarray}
	r_{exact} = \frac{\mathcal{P}_{h}(k_{*t})|_{freeze-out}}{\mathcal{P}_{\zeta}(k_{*s})|_{freeze-out}}. \label{rexact}
\end{eqnarray}
A common approximation is to calculate $r$ at sound horizon crossing ($c_sk=aH)$, assuming that the tensor and scalar modes freeze out at roughly the same time. To second order, this expression is given by \cite{Kinney:2007ag},
\begin{eqnarray}
	r_{approx} = 16c_{s}\epsilon \left[ 1+2\kappa - b(\eta +\kappa) \right].
\label{rapprox}
\end{eqnarray}
The approximation above, however, is only valid when scalar and tensor modes cross the horizon at similar instants \cite{Garriga:1999vw}. Since we keep the speed of sound $c_{s}$ and its dynamical evolution general, we do not assume this apriori and instead calculate $r$ directly using the ratio $\mathcal{P}_{h} / \mathcal{P}_{\zeta}$, (\ref{rexact}), by solving the mode equations for $u_{k}(k=k_{*s})$ and $v_{k}(k=k_{*t})$, and calculating $\mathcal{P}_{h}$ and $\mathcal{P}_{\zeta}$ at freeze-out. 

To first order we can write an expression for $r_{exact}$ as,
\begin{eqnarray}
	r_{exact} & \approx & \frac{\mathcal{P}_{h}(k_{*t})}{\mathcal{P}_{\zeta}(k_{*s})} \\
	& = & \frac{\left[ 1-2\epsilon_{h} + 2b\epsilon_{h} \right]}{\left[ 1-2\epsilon_{\zeta} + 2b \left( \epsilon_{\zeta}+\frac{\eta_{\zeta}}{2}-\kappa_{\zeta} \right) \right]} 16c_{s}\epsilon_{\zeta} \left(\frac{H_{h}}{H_{\zeta}}\right)^{2}, \nonumber \\
\end{eqnarray}
where the approximation above means that we have assumed instantaneous freeze-out of the scalar and tensor power spectra at their respective horizon crossings. Here $\epsilon_{h}$ and $H_{h}$ are calculated at $k_{*t}=aH$, and $c_{s}$, $\epsilon_{\zeta}$, $\eta_{\zeta}$ and $\kappa_{\zeta}$ are calculated at $c_{s}k_{*s}=aH$. Now for $c_{s}(k_{*s}) < 1$, scalar modes leave the horizon at an earlier time compared to the tensor modes. So for $c_{s}(k_{*s}) \ll 1$ we expect $H_{h}<H_{\zeta}$, and since $(H_{h}/H_{\zeta})^{2}$ is a stronger effect than the single factor of $\epsilon_{\zeta}$, we expect therefore $r_{exact}/r_{approx}<1$. Similarly for $c_{s}(k_{*s}) \gg 1$ we expect to get $r_{exact}/r_{approx}>1$. We verify these results numerically in Sec. \ref{Foura}, and find that the approximate expression for $r$ can give significant discrepancies from the actual tensor to scalar ratio for models in which $c_s\neq 1$.

This behavior was recently shown to give a modified expression for the tensor-to-scalar ratio \cite{Lorenz:2008et}, 
\begin{eqnarray}
	r = 16c_{s}\epsilon \left[ 1 + 2\kappa - b(\eta+\kappa) + 2\epsilon \ln c_{s} \right],
\label{rhor}
\end{eqnarray}
which we find is an excellent analytical approximation for our $r_{exact}$. The fact that $r_{exact}$ differs significantly from $r_{approx}$ tells us that, as $c_{s}$ deviates from 1, the next-to-leading order contribution in the expression for $r$ becomes important, the Taylor expansion which is often used breaks down, and next order terms, as given in (\ref{rhor}), are required.

%==========================================================================
\begin{figure*}[t]
\centering{
{\includegraphics[height=2.63in]{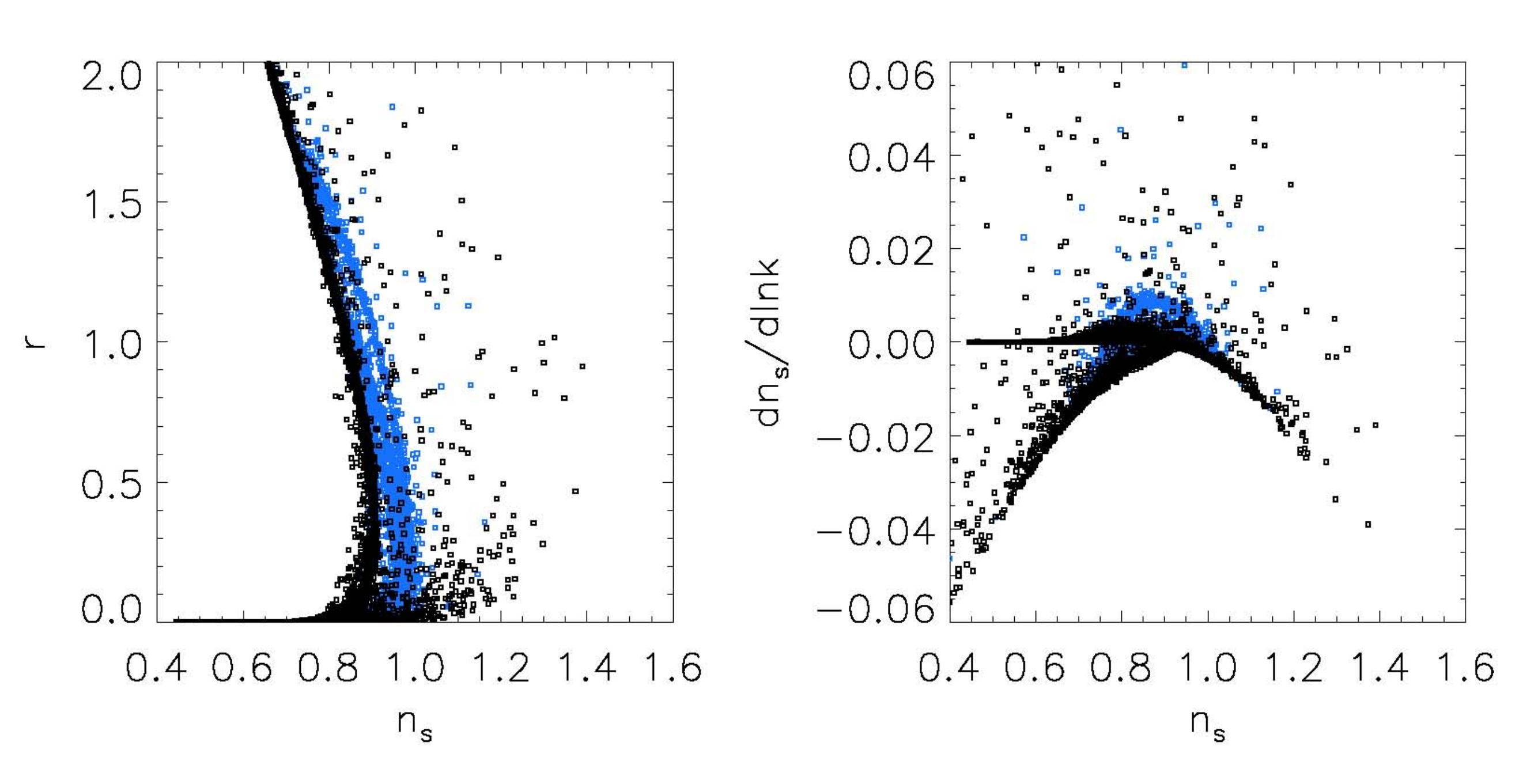}}
{\includegraphics[height=2.5in]{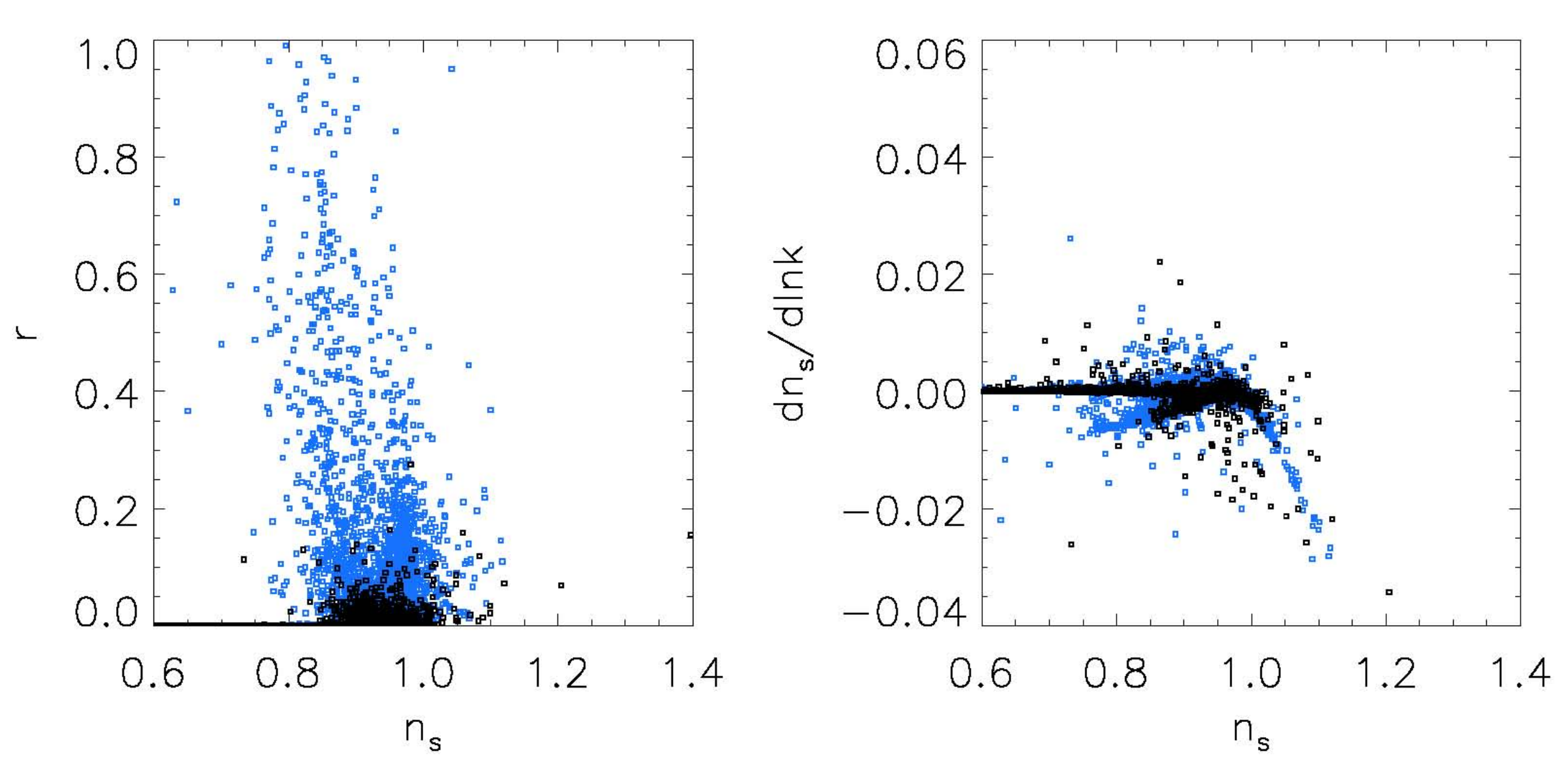}}
\caption{Monte Carlo sampling of canonical ($c_s=1$, $\kappa={^l \alpha}=0$) [upper panels] and general [lower panels] inflation models plotted in the $(r,n_{s})$ (left) and $(n_{run},n_{s})$ (right) planes, truncated at $l=5$ in the $^l\lambda$ and  $^l\alpha$ flow hierarchies using linear priors (black points) and log priors (blue points) on $a(t_{en})$, $c_{s}$, and flow parameters, in the ranges specified in (\ref{MCgeneral}). For general inflation we plot models that have $c_{s}(k_{*s}) \in [0,1]$.}
\label{fig1}}
\end{figure*}
%==========================================================================

%%%%%%%%%%%%%%%%%%%%%%%%%%%%%%%%%%%%%%%%%%%%%%
\section{Observable predictions of inflation}
\label{Four}
%%%%%%%%%%%%%%%%%%%%%%%%%%%%%%%%%%%%%%%%%%%%%%

We apply the formalism outlined in Secs. \ref{Two} and \ref{Three} to generate evolutionary trajectories for a general inflationary model. When constraining flow parameters with observational data typically two conditions can be considered:

{\it Condition 1:} Constraints on the flow parameters at horizon crossing from the form of the observed primordial power spectrum.

{\it Condition 2:} The end of inflation arises when $\epsilon=1$, when accelerated expansion as defined by (\ref{acceleq}) ceases, and requires that observable scales crossed the horizon a reasonable number of e-foldings, say, $N_e\sim 50-80$, before the end of inflation.

Monte Carlo Markov Chain analyses placing constraints on the flow parameters often solely impose Condition 1, e.g. \cite{Peiris:2006ug,Lesgourgues:2007aa,Hamann:2008pb,Lorenz:2008je}, while other analyses additionally impose the more restrictive, theoretically motivated, restriction in Condition 2, e.g. \cite{Kinney:2002qn,Easther:2002rw,Peiris:2006sj,Powell:2007gu,Peiris:2008be}.

%==========================================================================
\begin{figure*}[t]
\centering{
{\includegraphics[width=5in]{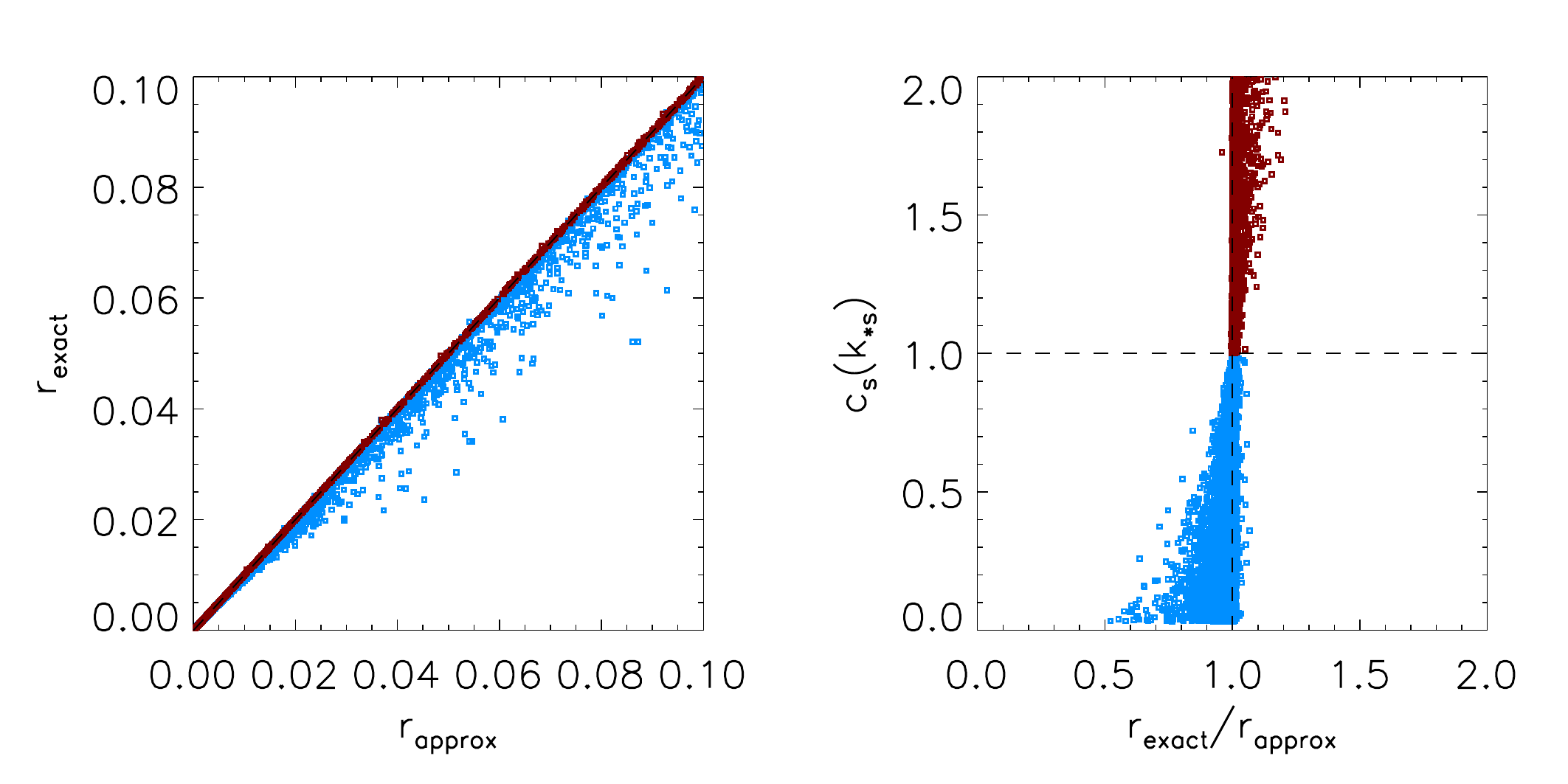}}
\caption{The difference between $r_{exact}$ and $r_{approx}$ [left panel] and the dependence on $c_s(k_{*s})$ [right panel] for general inflation models with $c_s(k_{*s}) \in [0,1]$ (blue points), and $c_{s}(k_{*s}) \in [1,2]$ (dark red points), for an order 5 Monte-Carlo simulation with ranges as given in (\ref{MCgeneral}).}
\label{fig2}}
\end{figure*} 
%==========================================================================
  
In Sec. \ref{Foura} we consider the properties of inflationary evolutionary trajectories, and the resultant power spectra, under Condition 2. In Sec. \ref{Fourb} we apply constraints on the flow parameters from the current WMAP 5-year data, Sloan Digital Sky Survey (SDSS) Luminous Red Galaxies (LRG) galaxy power spectrum and ``Union'' Type 1a supernovae data sets using Condition 1, and additionally consider the permitted models under the more restrictive Condition 2. 

%%%%%%%%%%%%%%%%%%%%%%%%%%%%%%%%%%%%%%%%%%%%%%
\subsection{Monte Carlo simulations of inflationary trajectories}
\label{Foura}
%%%%%%%%%%%%%%%%%%%%%%%%%%%%%%%%%%%%%%%%%%%%%%

%==========================================================================
\begin{figure*}[t]
\centering{
{\includegraphics[width=6in]{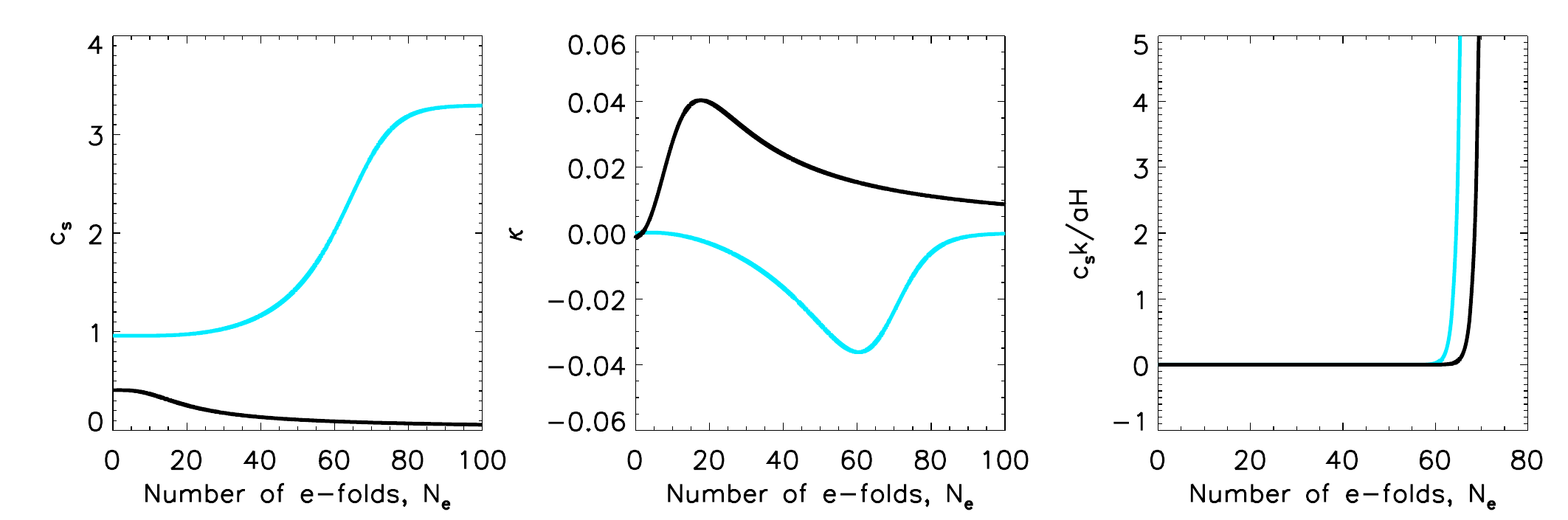}}
\caption{Evolution of $c_s$ [left panel], $\kappa$ [center panel] and $c_{s}k/aH$ [right panel] for $k=0.01$ Mpc$^{-1}$, for two example trajectories with $\kappa>0$ (black) and $\kappa<0$ (light blue) at the epoch when current cosmological scales exit the horizon. We find that for all viable trajectories, $c_sk/aH$ is monotonically decreasing over the course of inflation so that one does not need to be concerned about the prospect of multiple horizon crossings, and freezing and thawing of scalar perturbations.}
\label{fig3}
}
\end{figure*} 
%==========================================================================

We randomly select values at the end of inflation (when $\epsilon=1$) of the flow parameters $\{\kappa,{^l\lambda},{^l\alpha}\}$, the scale factor $a(t_{en})$ and the sound speed, $c_s$, within the following intervals:
\begin{eqnarray}
	a(t_{en}) & \in & [5 \times 10^{-29}, 5 \times 10^{-26}] , \nonumber \\
	c_{s} & \in & [0,2], \nonumber \\
	\kappa & \in & [-0.05, 0.05], \nonumber \\
	^{l}\lambda, ^{l}\alpha
	& & \left\{ 
		\begin{array}{lll}  
		\in & [-0.05, 0.05] \ \ \ & (l\le 5), \\
		= & 0\ & (l> 5).
	\end{array}
	\right.
\label{MCgeneral}
\end{eqnarray}
The Hubble constant at this instant is \cite{Bean:2007hc},
\begin{eqnarray}
	H(t_{en}) = \left( \frac{3.17708 \times 10^{-30}}{a(t_{en})} \right)^{2} M_{pl}.
\end{eqnarray} 

We also consider models of inflation that allow superluminal propagation of density perturbations, with the speed of sound at the end of inflation, $c_{s} \in [0,2]$. Faster-than-light propagation has been shown to arise in higher order QCD corrections, see for example \cite{Drummond:1979pp,Daniels:1995yw,Scharnhorst:1990sr}, and in many other theories, such as \cite{Hashimoto:2000ys,Nastase:2006na,Jacobson:2004ts}. It has been shown recently \cite{Babichev:2007dw} that superluminal propagation in generic k-$essence$ theories does not lead to the appearance of closed causal curves (hence they do not violate causality).

To obtain the scalar perturbation spectrum we evolve the parameters back to the time, $N_{\zeta i}$ when $y_\zeta/(1-\epsilon-\kappa)|_{N_{\zeta i}}=50$, and then evolve $u_k$  forward to freeze-out of the power spectrum. The pivot mode crosses the horizon at $N_e=N_{*}$, where $k_{*s} = aH/c_s|_{N_{*}}$. If $N_{*}\in [50,80]$ then the observable parameters $\{A_s,n_s,dn_s/d\ln k\}$  are calculated and the trial is recorded. The tensor spectrum, and $r$, are calculated in an analogous manner, by evolving back to $N_{h i}$, when $y_h/(1-\epsilon)|_{N_{h i}}=50$, and evolving $v_k$ forward  to obtain ${\cal P}_{h}(k_{*t})$.

In Fig. \ref{fig1} we constrast the properties of inflationary trajectories satisfying Condition 2 in the case of canonical inflation ($c_s=1,$ $\kappa= {^l\alpha}=0$), as discussed in \cite{Kinney:2003uw,Peiris:2003ff}, and in models of general inflation. The introduction of an evolving sound speed noticeably alters the distribution of spectrum observables arising from the flow trajectories. The asymptotic relation $n_s-1 \approx -r/8$ that holds for canonical inflation is broadened to $n_s-1 \approx -r/8c_s$, and the introduction of a non-zero $\kappa$ gives rise to nearly scale-invariant models with non-zero running. Allowing superluminal propagation, with $c_s>1$, can give scenarios with larger tensor-to-scalar ratios \cite{Mukhanov:2005bu}.

For the main analysis in Sec. \ref{Four}, we assume linear priors on the flow parameters (most consistent with assuming linear priors on the power spectrum observables at lowest order). We note, for interest, however that Monte Carlo sampling assuming log priors on the flow parameters can alter the sampling of allowed models, as  in the context of canonical inflation \cite{Peiris:2008be,Valkenburg:2008cz}. The effect is more noticeable in general inflationary models where introducing log priors can allow larger tensor scenarios to be sampled more efficiently. 

In Fig. \ref{fig2} we demonstrate the difference between the exact tensor-to-scalar ratio coming from fully evolving both ${\cal P}_{\zeta}$ and ${\cal P}_{h}$, (\ref{rexact}), and well approximated by (\ref{rhor}), and the ratio derived by assuming both tensor and scalar freeze-out concurrently at  scalar horizon crossing (\ref{rapprox}). As introduced in Sec. \ref{Threec}, for sound speeds different from 1, and especially as $c_s \rightarrow 0$, the discrepancy between the two values becomes significant, as much as 50-60\%.  We can understand these deviations by the fact that as $c_s$ deviates from 1, tensor and scalar modes leave the horizon, and are frozen, at increasingly disparate epochs.

Since the tilt $n_{s}$ and running $n_{run}$ are calculated purely from the scalar power spectrum (with no reference to the tensor power spectrum), we expect that the values of corresponding $n_{s,exact}$ and $n_{run,exact}$, found directly from the power spectrum, will be similar to the values obtained from the numerical expressions (\ref{nsapprox}) and (\ref{nrunapprox}). We have verified numerically that the approximate expression for $n_{s}$ is as good as the exact calculation to within a few percent.  The fact that $n_{s,exact}$ and $n_{s,approx}$ are in good agreement also implies that our approximate expression for $n_{s}$ to second order is reasonable, and we do not need to calcualte $n_{s}$ to fourth or fifth order.

As shown in Fig. \ref{fig3}, at the epoch when observable modes cross the horizon, $\kappa$ and $c_s$ may be increasing or decreasing, $c_sk/aH$, however, always decreases monotonically. Therefore once the modes have left the horizon (i.e. $c_{s}k/aH < 1$) they do not re-enter and we do not have to worry about the presence of multiple horizon crossings (with the potential for unfreezing and refreezing of fluctuations). 

%%%%%%%%%%%%%%%%%%%%%%%%%%%%
\subsection{Constraints from cosmological observational data}
\label{Fourb}
%%%%%%%%%%%%%%%%%%%%%%%%%%%%

We have included our general inflationary perturbation code into the CAMB code \cite{Lewis:1999bs} to evolve background equations and first order density perturbations for a flat universe containing baryons, CDM, radiation, massless neutrinos and use CosmoMC \cite{Lewis:2002ah} to perform a Monte-Carlo-Markov-Chain analysis of the model parameter space in comparison to current cosmological data. 

In Table \ref{Table1} we summarize the priors on the flow parameters for five inflationary scenarios we investigate. We use linear priors on the flow parameters, $\{\epsilon,\kappa,{^l\lambda},{^l\alpha}\}$, up to some $l_{max}$ for ${^l\lambda}$ and ${^l\alpha}$, $c_s$  and $\ln (10^{10} A_s)$ at horizon crossing for $k_{*s}=0.01$ Mpc$^{-1}$. $A_s$ is used to calculate the value of $H(N_{*})$ at horizon crossing using (\ref{Pzeta}) and (\ref{normalization}), which then gives $a(N_{*})=c_sk_{*s}/H|_{N_{*}}$.  We choose truncations at $l=2$ (C1) and $l=5$ (C2) to demonstrate the effect of adding in extra degrees of freedom in reconstructing the power spectrum. 

%==========================================================================
\begin{table*}[t!]
\begin{center}
\begin{tabular}{|c|c|c|c|c|c|c|c|c|}
\hline
Scenario & Inflation type & $c_s$ & $\epsilon$ & ${^{l}\lambda}$  & $\kappa$ & ${^{l}\alpha}$   & $\Delta (-2\ln L)$ & $\Delta (d.o.f.)$
\\
\hline
C1 & Canonical  & 1     & [0,0.5] & [-0.1,0.1], $l_{max}=2$ & 0          & 0 & 1.14 & 0 \\
C2 & Canonical  & 1     & [0,0.5] & [-0.5,0.5], $l_{max}=5$ & 0          & 0 & 1.18 & 3 \\
G1 & General    & [0,1] & [0,0.5] & [-0.5,0.5], $l_{max}=2$ & [-0.5,0.5] & 0 & 1.16 & 2 \\
G2 & General    & [0,2] & [0,0.5] & [-0.5,0.5], $l_{max}=2$ & [-0.5,0.5] & 0 & 0.96 & 2 \\
G3 & General    & [0,2] & [0,0.5] & [-0.5,0.5], $l_{max}=2$ & [-0.5,0.5] & [-1.0,1.0], $l_{max}=1$ & 1.28 & 3 \\
\hline
\end{tabular}
\end{center}
\caption{Summary of the parameter ranges investigated for each of the inflationary scenarios in the MCMC analysis in Sec. \ref{Fourb}. All ranges are for values as the scalar mode $k_{*s}$ crosses the horizon. We also show the change in the effective minimum $\chi^2=-2\ln L$, where $L$ is the likelihood, and number of extra degrees of freedom (d.o.f.) in comparison to the fiducial, canonical power law primordial power spectrum with scale independent running.}
\label{Table1}
\end{table*}
%==========================================================================

Using the approach described in Sec. \ref{Three} we calculate the scalar and tensor power spectra for $5\times 10^{-6}$ Mpc$^{-1} \le k \le 5$ Mpc$^{-1}$. For the MCMC analysis, we purely consider constraints within this range of observable scales, i.e. Condition 1 of Sec. \ref{Four}. We do not impose the stricter requirement of Condition 2, that pertains to the full inflationary history, however we do require that $0\le \epsilon< 1$ (and hence that inflation persists) during all times from when the initial conditions are set, up to when the power spectrum has converged for all observable $k$-modes. We discuss the effect that this additional condition has on the parameter constraints below.

%==========================================================================
\begin{figure*}[t]
\centering{
{\includegraphics[width=5.5in]{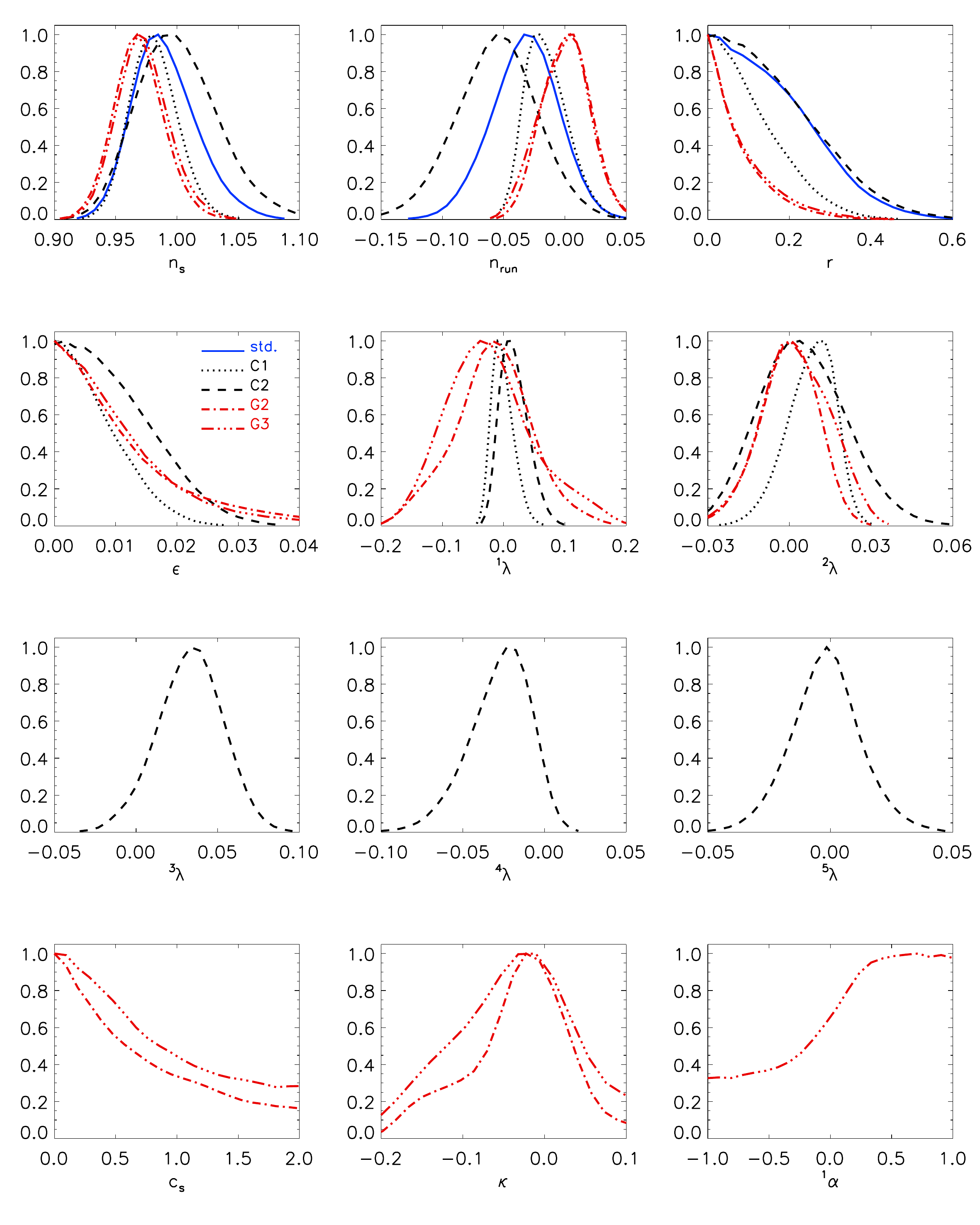}}
\caption{Comparison of the 1D marginalized posterior probability distributions for the flow parameters and observables, for the cases of power law inflation with scale-independent running (blue full line) and our models C1 (black dotted), C2 (black dashed), G2 (red dot-dashed) and G3 (red triple dot-dashed). The constraints for model G1 are the same as those for G2, with just $c_s$ cutting off at 1, so we do not show them here.}
\label{fig4}
}
\end{figure*} 
%==========================================================================

We constrain the models using a combination of cosmological datasets, including measurements of the CMB temperature and polarization power spectrum from the WMAP 5-year data release \cite{Nolta:2008ih,Dunkley:2008ie}, the ``Union'' set of supernovae \cite{Kowalski:2008ez}, and the matter power spectrum of Luminous Red Galaxies (LRG) as measured by the Sloan Digital Sky Survey (SDSS) \cite{Tegmark:2006az, Percival:2006gt}. We include the shift parameter, $a_{scl}$, to adjust the matter power spectrum as discussed in \cite{Tegmark:2006az}.

The MCMC convergence diagnostic tests on each scenario considered are performed on $4$ or more chains using the Gelman and Rubin ``variance of chain mean''$/$``mean of chain variances'' $R$ statistic for each parameter. Our 1D and 2D constraints are obtained after marginalization over the remaining ``nuisance'' parameters, again using the programs included in the CosmoMC package. 
 
\begin{figure*}[t]
\centering{
{\includegraphics[width=2.25in]{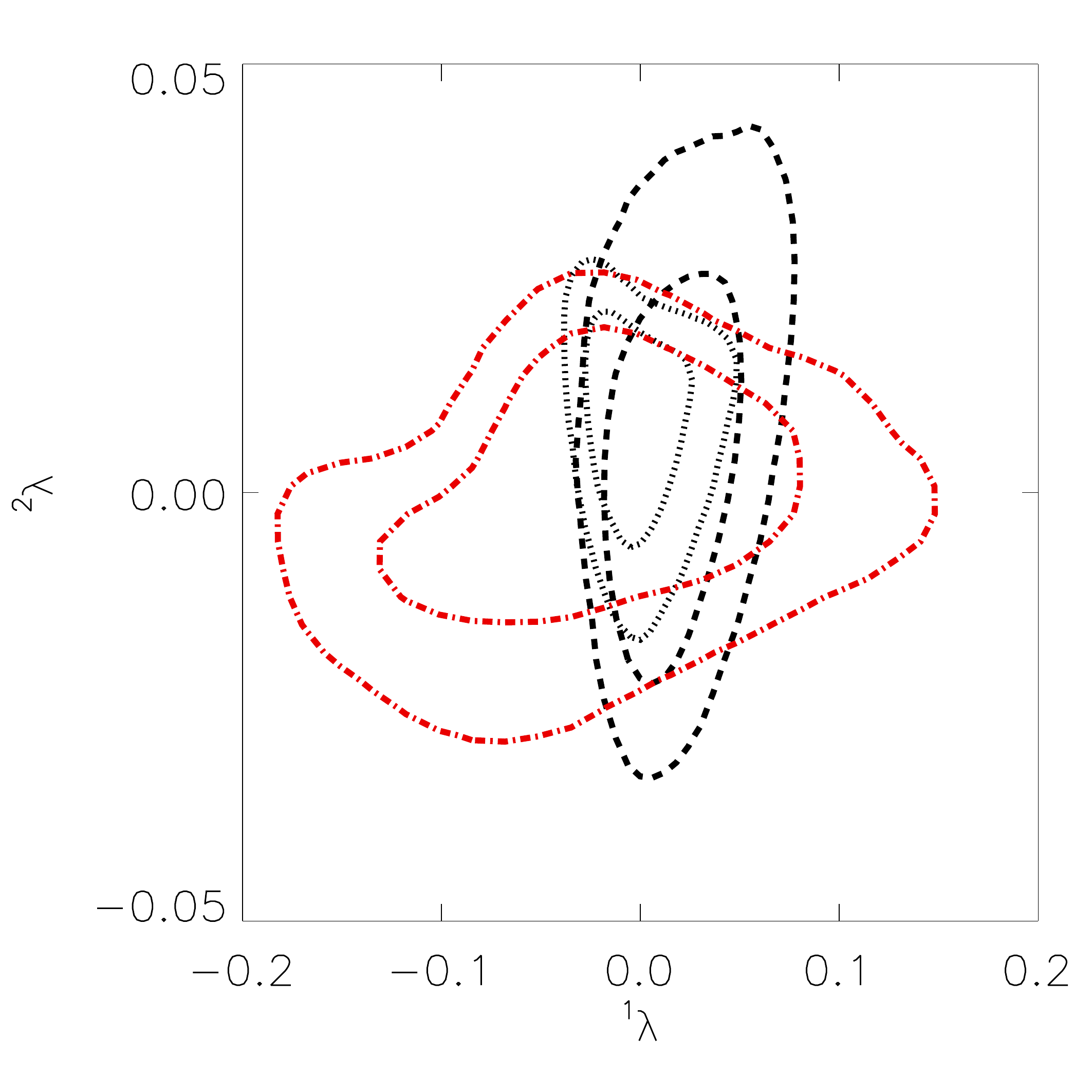}
\includegraphics[width=2.25in]{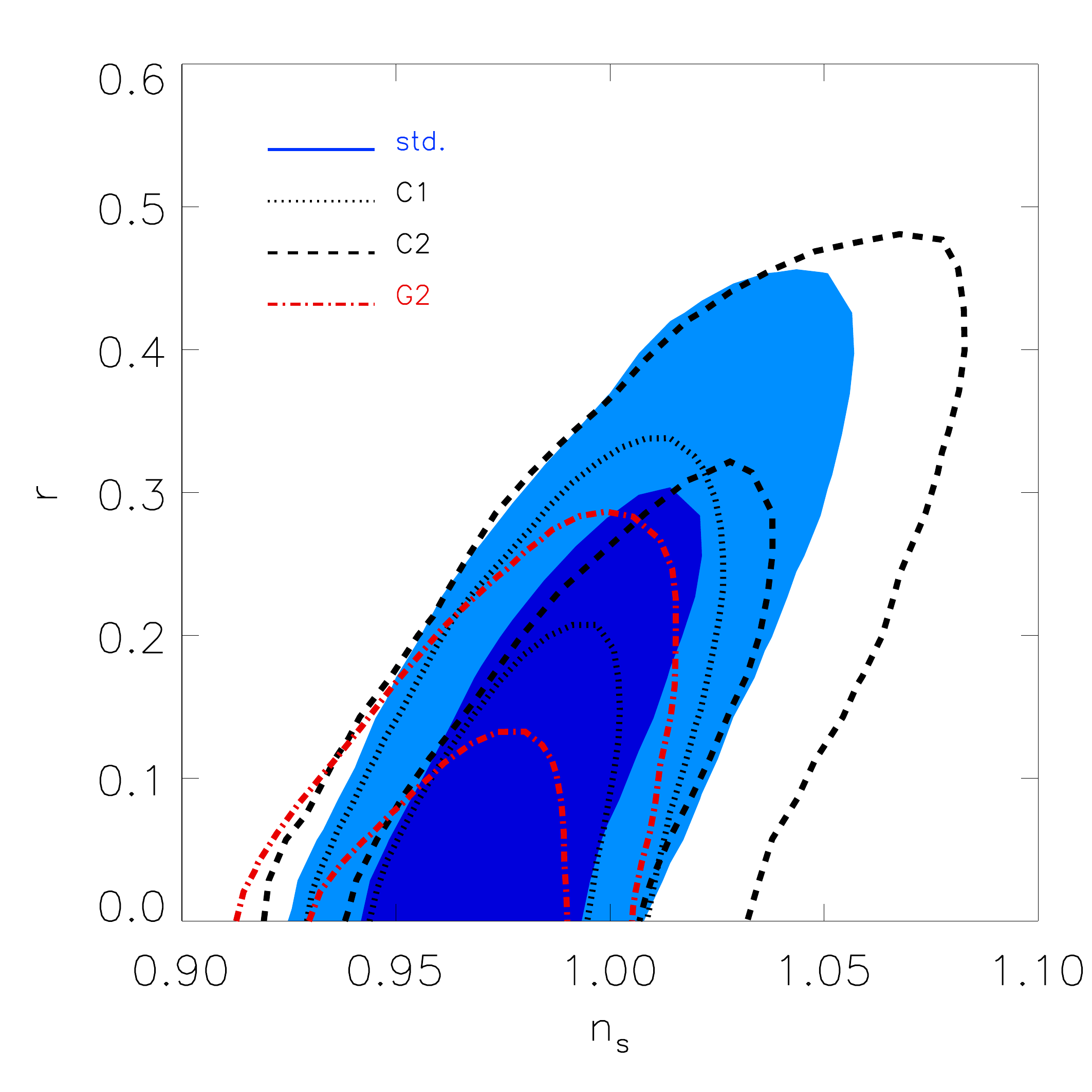}
\includegraphics[width=2.25in]{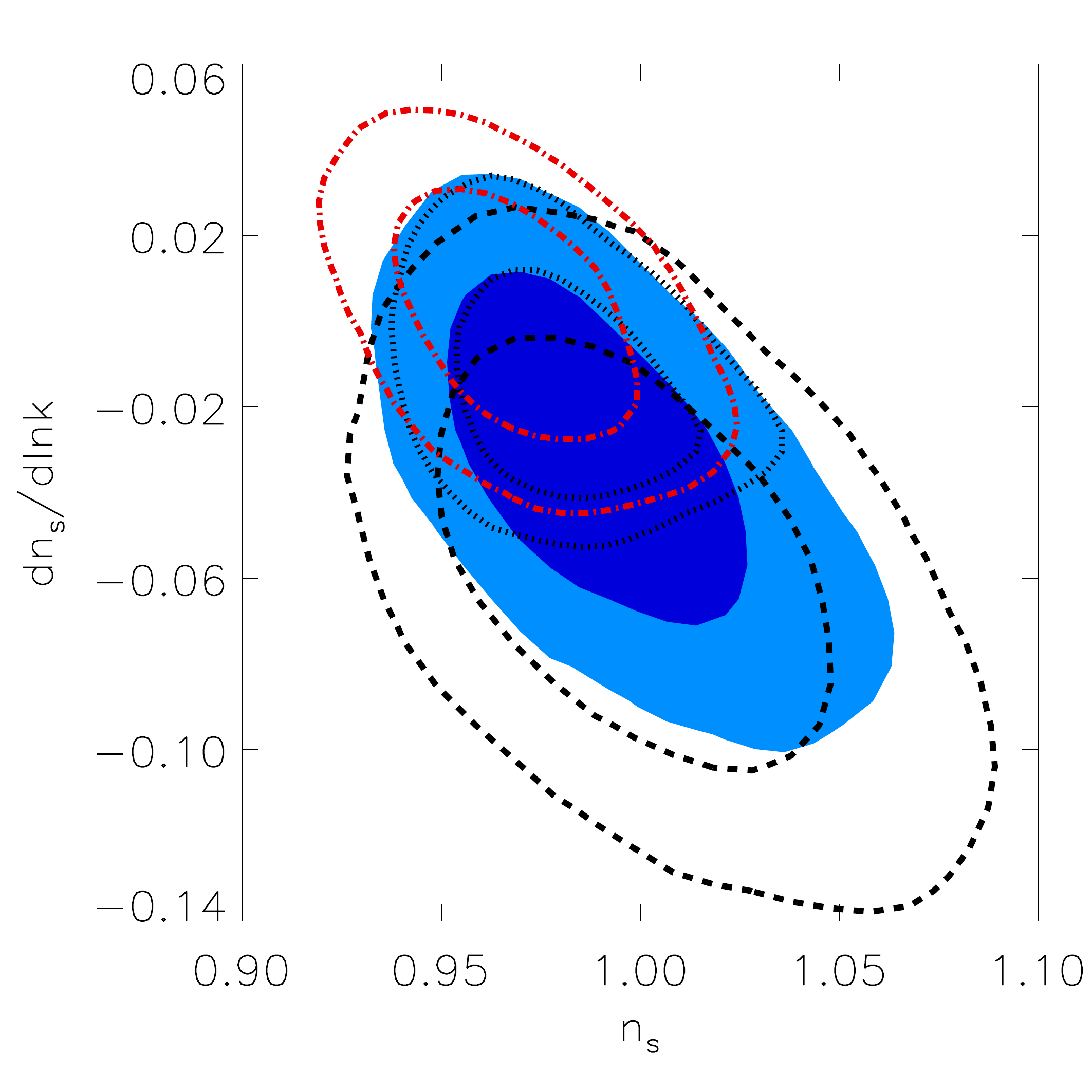}
}
\caption{The 68\% and 95\% confidence regions of the $(^1\lambda,^2\lambda)$ [left], $(r,n_{s})$ [center], and $(n_{run},n_{s})$ [right] parameter spaces. Constraints for a standard power law spectrum with constant running are shown in dark blue and light blue, as well as canonical inflation models C1 (black dotted), C2 (black dashed) and general inflation model G2 (red dot-dashed).}
\label{fig5}}
\end{figure*}

In Figs. \ref{fig4} and \ref{fig5} we show the 1D and 2D marginalized posterior probability distributions for the flow parameters and power spectrum observables, $n_s$, $n_{run}$, and $r$ at $k_{*s}$ for the canonical and general inflationary scenarios studied, and for the ``standard'' power law inflationary model with scale independent running. Comparing the constraints on the canonical and general models, we see that inclusion of higher order flow parameters can noticeably change the constraints on the power spectrum properties.
%Comparing the constraints on models C1 and C2, we see that inclusion of higher order flow parameters can noticeably change the constraints on the power spectrum properties. Model C2, which has a larger number of flow parameters can allow larger negative running models and large tensor amplitudes comparable with the power law model. This is consistent with our intuition that the most accurate description should contain an infinite number of higher order derivatives of the potential or Hubble parameter. One also finds that the data does not require higher order flow parameters to be necessarily negligible compared to lower order parameters. These higher order parameters are however constrained by the data:  increasing the magnitudes of the higher order flow parameters increases the variation of the scalar spectral index over the observed scales and significantly boosts or diminishes small and large scale power to levels inconsistent with CMB and galaxy matter power spectrum observations respectively. 
%Including higher order flow parameters opens up the allowed parameter space for lower order flow parameters. 
%In Figure \ref{fig5}, we plot the 2D marginalized likelihoods in the $({^1\lambda},{^2\lambda})$, $(r,n_{s})$ and $({n_{run}},n_{s})$ planes contrasting  our models C1, C2, and G2 against a power law inflationary power spectrum with scale independent running of the spectral index. 
Model C2 significantly opens up the accessible region as compared to the standard power law model and C1, allowing larger negative running models and large tensor amplitudes. That one obtains constraints on all higher order flow parameters ${^3\lambda}-{^5\lambda}$ separately in C2, can be attributed to our truncation of the flow hierarchy at $l=5$.  If one were to further include more higher order parameters in the analysis, then this would further open up the parameter space for lower-order parameters. 

Note that since the data has only a finite amount of information, adding extra degrees of freedom does not necessarily lead to a statisticaly important improvement in the fit. Table \ref{Table1} shows a comparison of the fit obtained for primordial power spectra based on the flow parameters for different models, with that for the commonly assumed canonical power law spectrum with scale independent running of the scalar spectral index. The alternative parameterization of the primordial spectrum does not  significantly improve the fit with data, the improvement in $\chi^2$ does not outweigh the additional degrees of freedom added, but rather allows the primordial power spectrum to be reconstructed with more freedom.

\begin{figure*}[t]
\centering{
\includegraphics[width=7in]{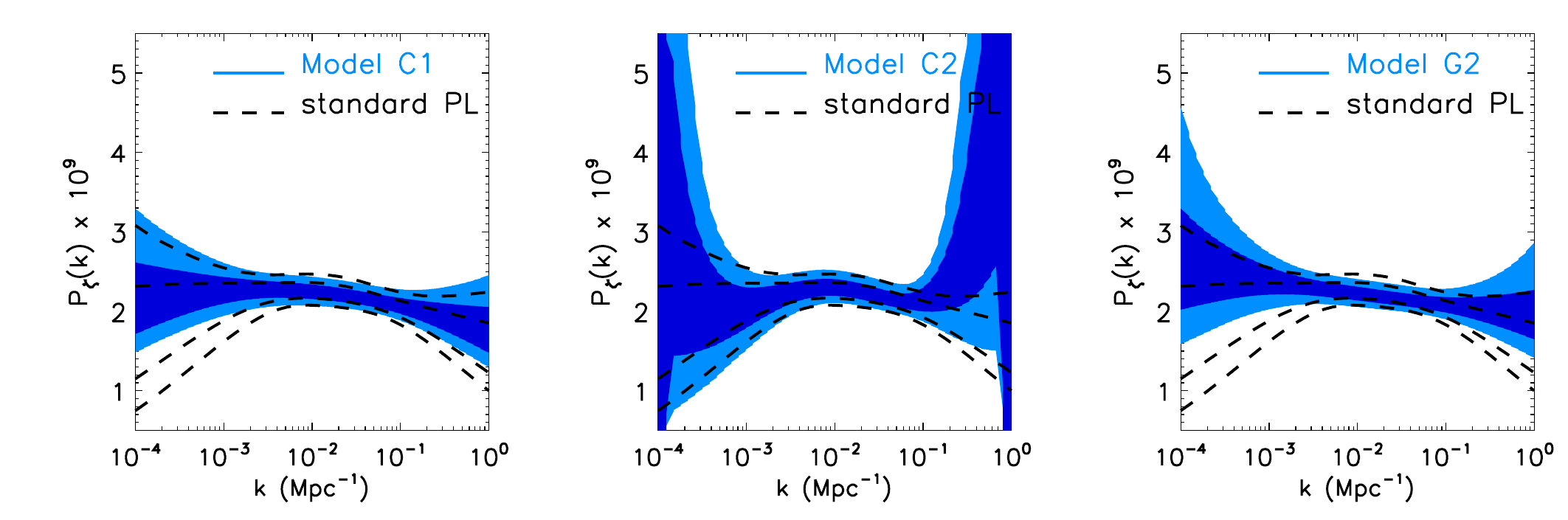}
\caption{Comparison of the 1D marginalized posterior probability distributions, showing 68\% (dark blue) and 95\% (pale blue) confidence limits, for the primordial scalar power spectrum for models C1 [left], C2 [center] and G2 [right], in comparison to a ``standard'' power law spectrum with scale independent running (black dashed lines). Increasing the number of flow parameters increases the freedom with which the spectrum is reconstructed. In particular we can get significantly greater or smaller power at small and large scales, least well measured by the CMB and large scale structure data respectively. The bounds on the higher-order parameters in model C2 directly arise from constraining this greater or lesser power at the extreme ends of the observed scales.}
\label{fig6}
}
\end{figure*}

Increasing the magnitudes of the higher order flow parameters increases the variation of the scalar spectral index over the observed scales and significantly boosts or diminishes small- and large-scale power to levels inconsistent with CMB and galaxy matter power spectrum observations respectively. We show this in Fig. \ref{fig6} where we plot the 1D posterior probability distributions for the primordial power spectrum for models C1, C2, and G2. Specifically, C2 is better able to fit freedom in the power spectrum at large and small scales arising from larger experimental and cosmic variance errors on those scales. We also find that notable degeneracies exist between the higher-order flow parameters in C2, reflecting that the number of independent degrees of freedom measured by the observed primordial power spectrum is less than the number of higher-order parameters employed. Future small-scale measurements of the power spectrum, for example using Lyman-$\alpha$ observations, will help to reduce this interdependency. 
%Note that here we are not contradicting what we had stated in \ref{Foura}, that canonical models do not favor large running scale invariant models, since here we have not imposed Condition 2 that $\epsilon=1$ at the end of inflation; we are working under Condition 1. 

Allowing general inflation, with $c_s \ne 1$, $\kappa \ne 0$, as in models G1-G3, predominantly alters the tensor amplitude posterior distribution, consistent with the overlap of observational constraints and the Monte Carlo sampling using linear priors on the flow parameters in general models shown in Fig. \ref{fig1}. The scalar spectral index to first order is dependent on $2(-2\epsilon+{^1\lambda}-\kappa)$ so that allowing an evolving sound speed (with $\kappa\neq 0$) opens up the range of ${^1\lambda}$ that is consistent with observations in comparison to canonical models.  

\begin{figure*}[t]
\centering{
{\includegraphics[width=2.25in]{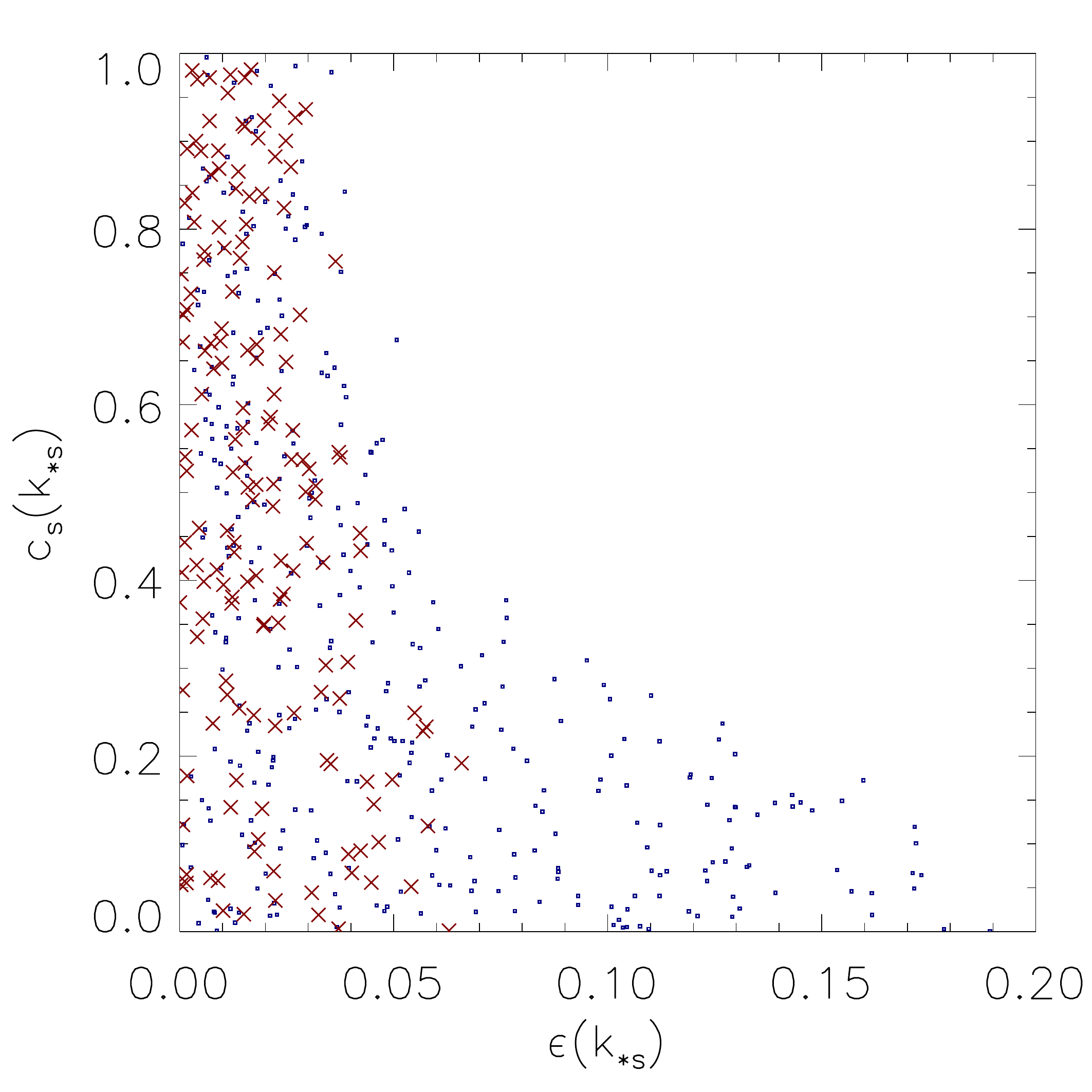}
\includegraphics[width=2.25in]{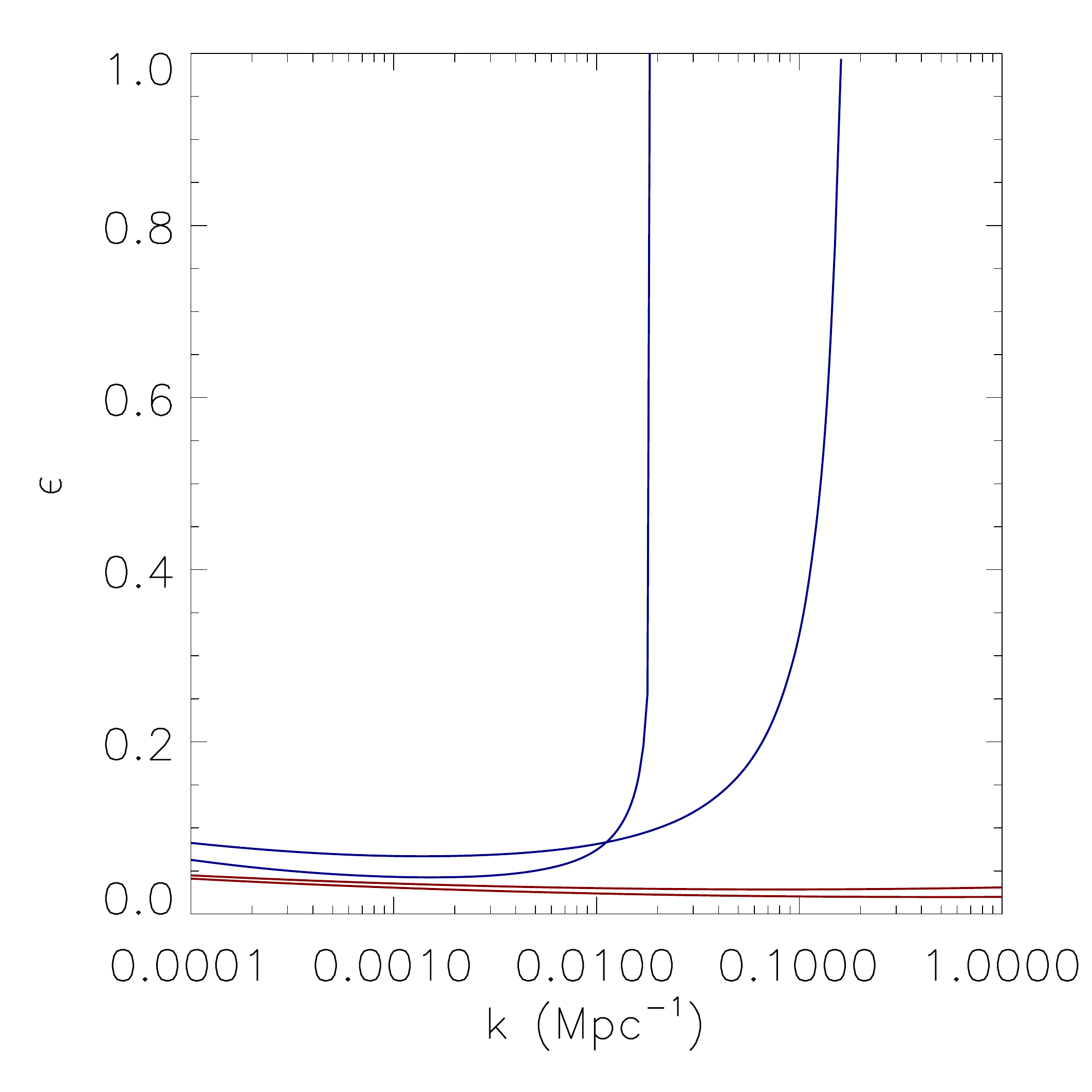}
}
\caption{While models with large $\epsilon$ and small $c_s$ are consistent with observational constraints from the bound on $r$,  imposing the condition that inflation persists while the observable scales exit the horizon places an additional restriction on $\epsilon(k_{*s})$. The left panel shows $c_{s}(k_{*s})$ vs. $\epsilon(k_{*s})$ for models which satisfy observational constraints. Models which additionally satisfy $\epsilon(k)< 1$ (red crosses) have an upper limit on the value of $\epsilon(k_{*s})$, while for larger $\epsilon(k_{*s})$ (blue squares) the $\epsilon(k)<1$ condition is not met (these models are rejected in the MCMC analysis). %and $\kappa(k_{*s})$
The right panel shows the evolution of $\epsilon(k)$ for observable modes, for example models which give constraints at the pivot point in agreement with the data, and which satisfy (red, lower two lines) or break (blue, upper two curves that reach $\epsilon=1$) the $\epsilon(k)<1$ constraint. }
\label{fig7}
}
\end{figure*}

The bounds on the flow parameters that we obtain in our analysis arise from two different sets of constraints: (i) observations, and (ii) the $\epsilon<1$ requirement.  In canonical models, we find the constraint on $\epsilon$ imposed by the $\epsilon<1$ condition is very similar to that arising from the observational constraint, therefore it does not play a major role. This is not true for general models, however. We see from (\ref{rapprox}) that the bound on $r$ alone places a rough upper bound on the product $c_s\epsilon$, but leaves $\epsilon$ and $c_s$ individually unbounded, as, for example, in \cite{Lorenz:2008je}. Insisting however, that, for consistency, inflation should occur over the observable scales in general inflationary models, i.e. $\epsilon<1$ from when the initial conditions for each observable mode are set up until when the power spectrum for all observable modes $10^{-4}$ Mpc$^{-1} \lesssim k \lesssim 5$ Mpc$^{-1}$ has converged, introduces a constraint on $\epsilon_N$ over the observable range. This restricts the value of $\epsilon(k_{*s})$ and $\kappa(k_{*s})$ over and above the observational constraints arising from the power spectrum properties at the pivot point. In Fig. \ref{fig7} we demonstrate this by plotting $c_{s}(k_{*s})$ vs. $\epsilon(k_{*s})$ for a sampling of flow parameter combinations for model G1 that are consistent with observations at the 95\% confidence level. We find that requiring $\epsilon<1$ induces the constraint on $\kappa(k_{*s})$ and constrains $\epsilon(k_{*s})\lesssim 0.03$ at the 95\% confidence level. Note that, as seen in Fig.~\ref{fig4}, adding in an extra flow parameter, ${^1\alpha}$, as in model G2, does not significantly alter the observational constraints on $\kappa$ in the presence of the $\epsilon<1$ requirement. Extending the truncation to include higher nonzero $^{l}\lambda$ might alter the constraints on $\epsilon$, however such additional degrees of freedom are not statistically warranted by the data, so we don't consider such models here.

\begin{figure*}[t]
\centering{
{\includegraphics[width=7in]{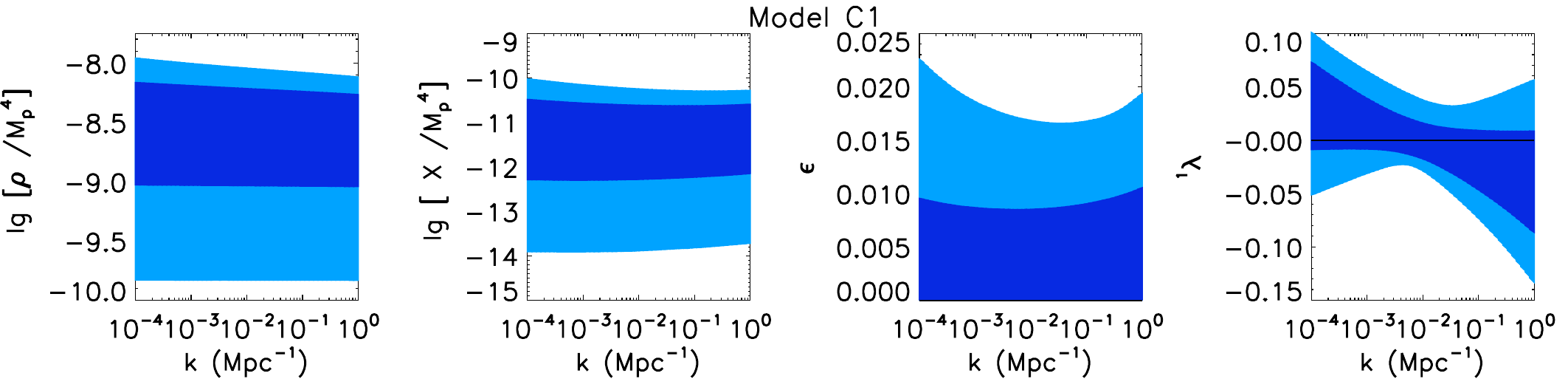}}
{\includegraphics[width=7in]{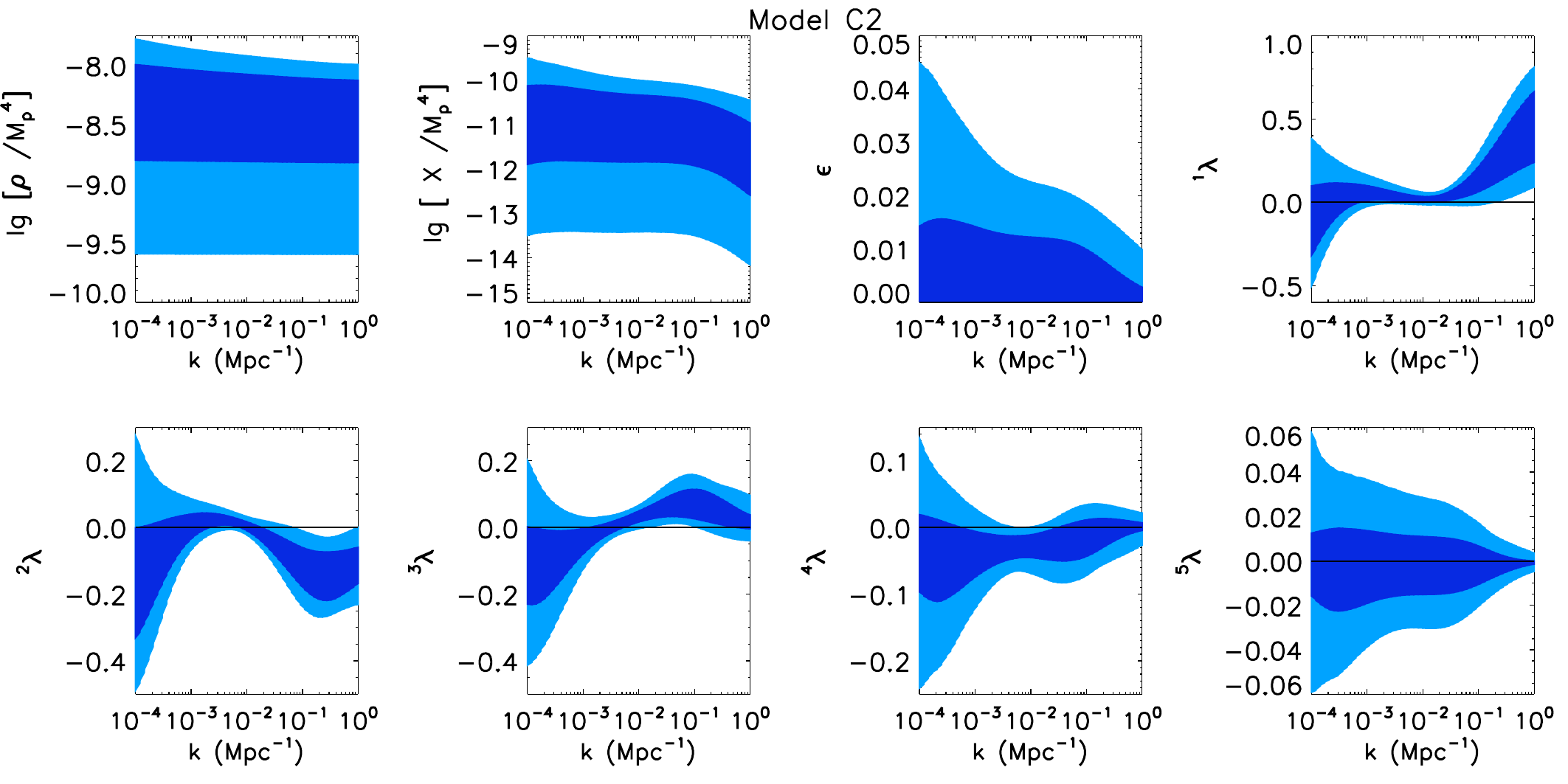}}
\caption{Comparison of the 1D marginalized posterior probability distributions  for Model C1 [top panels] and C2 [lower panels], showing 68\% (dark blue) and 95\% (pale blue) confidence limits, for the flow parameters and observables as each observed comoving mode $k$ exits the horizon, where $k(N)$ is given by (\ref{kN}).}
\label{fig8}
}
\end{figure*}

\begin{figure*}[t]
\centering{
{\includegraphics[width=6.in]{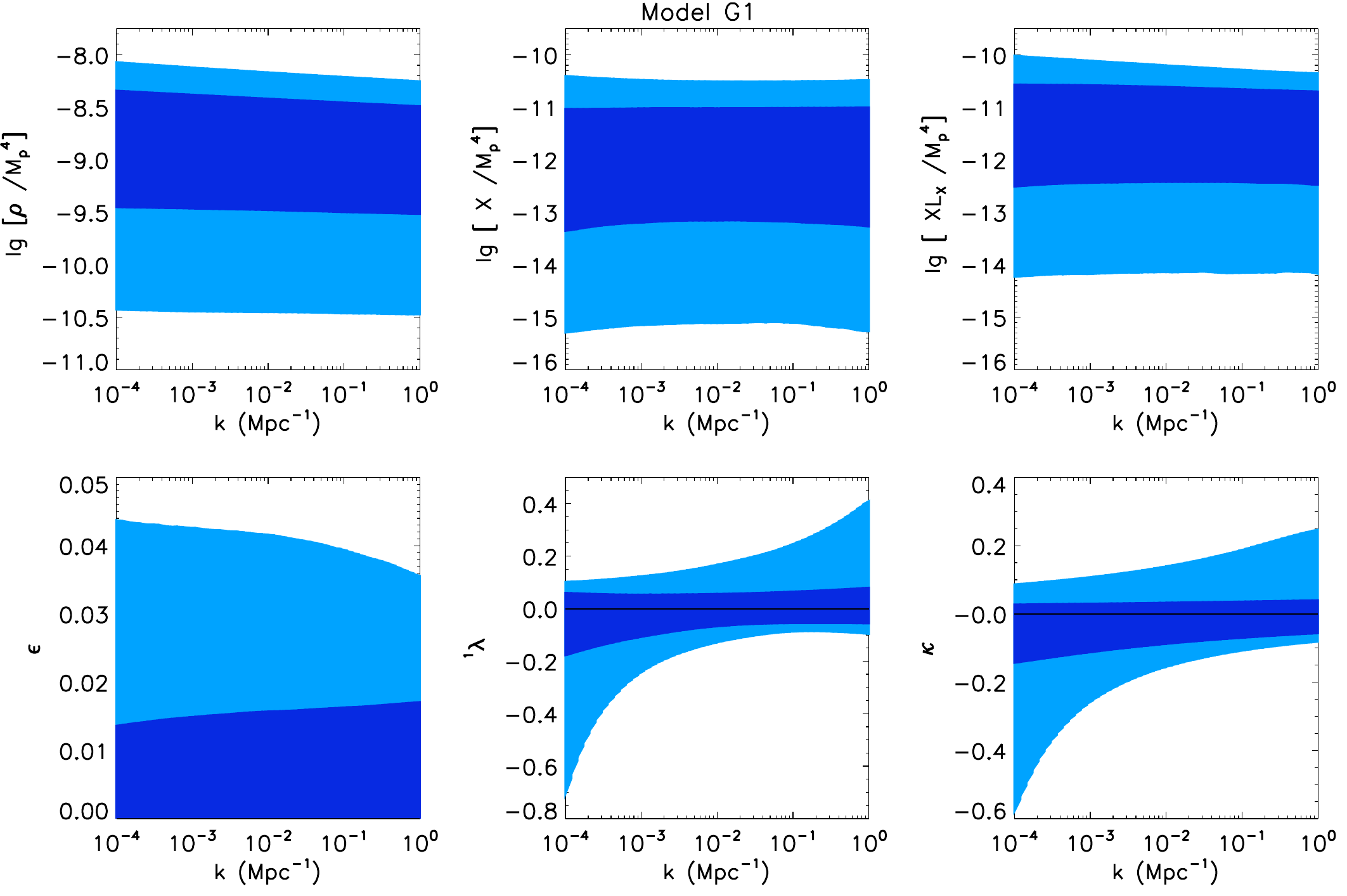}}
\caption{Comparison of the 1D marginalized posterior probability distributions for Model G1, showing 68\% (dark blue) and 95\% (pale blue) confidence limits,  for the flow parameters and observables as each observed comoving mode $k$ exits the horizon.}
\label{fig9}
}
\end{figure*}

In Figs. \ref{fig8} and \ref{fig9} we show the 1D 68\% and 95\% confidence levels for the flow parameters and $\rho$, $X$, and $X{\cal L}_X$, for the field choice ${\cal L}_X = c_s^{-1}$, as a function of the observable comoving mode $k(N)$, that exits the horizon $N$ e-folds before the end of inflation (the slice at $k_*=0.01$ Mpc$^{-1}$ is directly analogous with the 1D constraints shown in Fig. \ref{fig4}), where \cite{Bean:2008ga},
\bea
	\rho(N) & = & \rho(N_*)\exp\int_{N*}^{N}2\epsilon(N')dN', \\
	X{\cal L}_{X} (N) & = & \frac{1}{3}\epsilon(N)\rho(N), \\
	X(N) & = & \frac{1}{3}\epsilon(N)c_s(N)\rho(N), \\
	\ln k(N) & = & \ln k_{*s} -( N-N_{*}) -\int_{N_*}^{N}\epsilon(N')dN', \label{kN}\ \ 
\eea
with
\bea
	\rho(N_*) = \frac{24\pi^2c_s(N_*)\epsilon(N_*)A_s}{1-2\epsilon(N_*)+2b\left(2\epsilon(N_*)+\eta(N_*)-\frac{\kappa(N_*)}{2}\right)} \ \ 
\eea
As has been pointed out in previous analyses using power law primordial power spectra \cite{Cortes:2007ak}, the best measured modes are around $k \sim 0.01$ Mpc$^{-1}$, with large-scale constraints being limited by cosmic variance. However even in the scenarios with higher-order flow parameters allowed to vary, the observations, in combination with the $\epsilon <1$ condition, impose interesting constraints on the flow parameters across the full range of observable scales.

%==========================================================================
%==========================================================================

%==========================================================================
%==========================================================================

%==========================================================================
%==========================================================================

%==========================================================================
%==========================================================================

%==========================================================================
%==========================================================================

%==========================================================================
\begin{figure*}[t]
\centering{
{\includegraphics[width=6in]{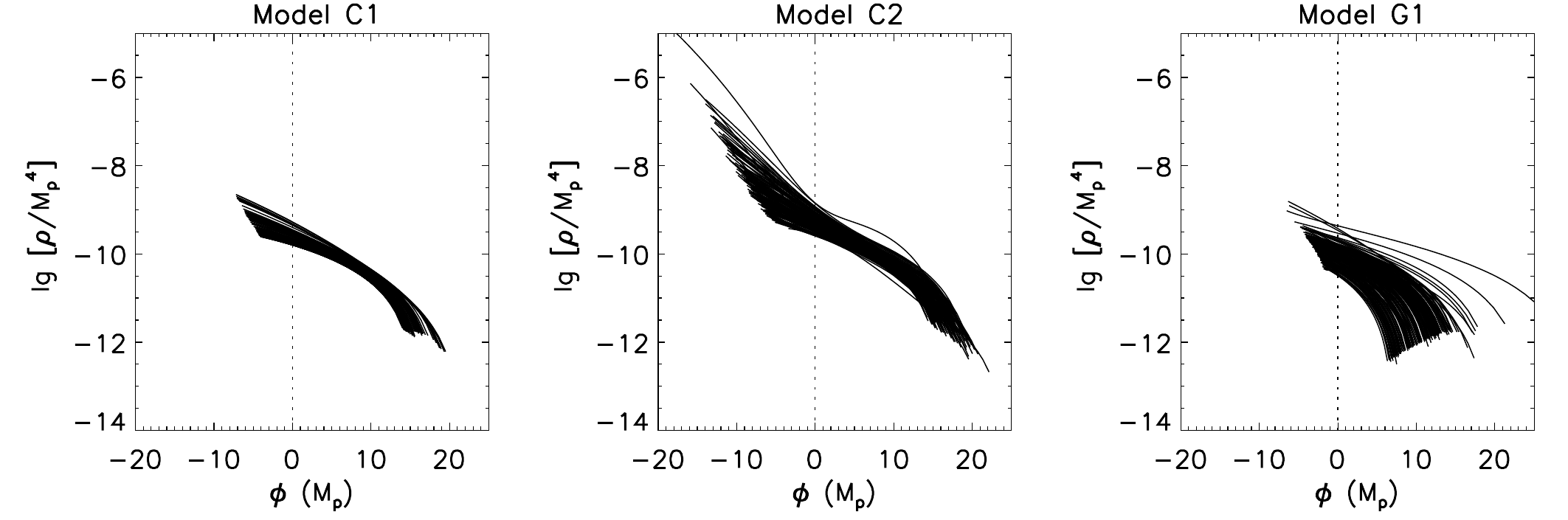}}
{\includegraphics[width=6in]{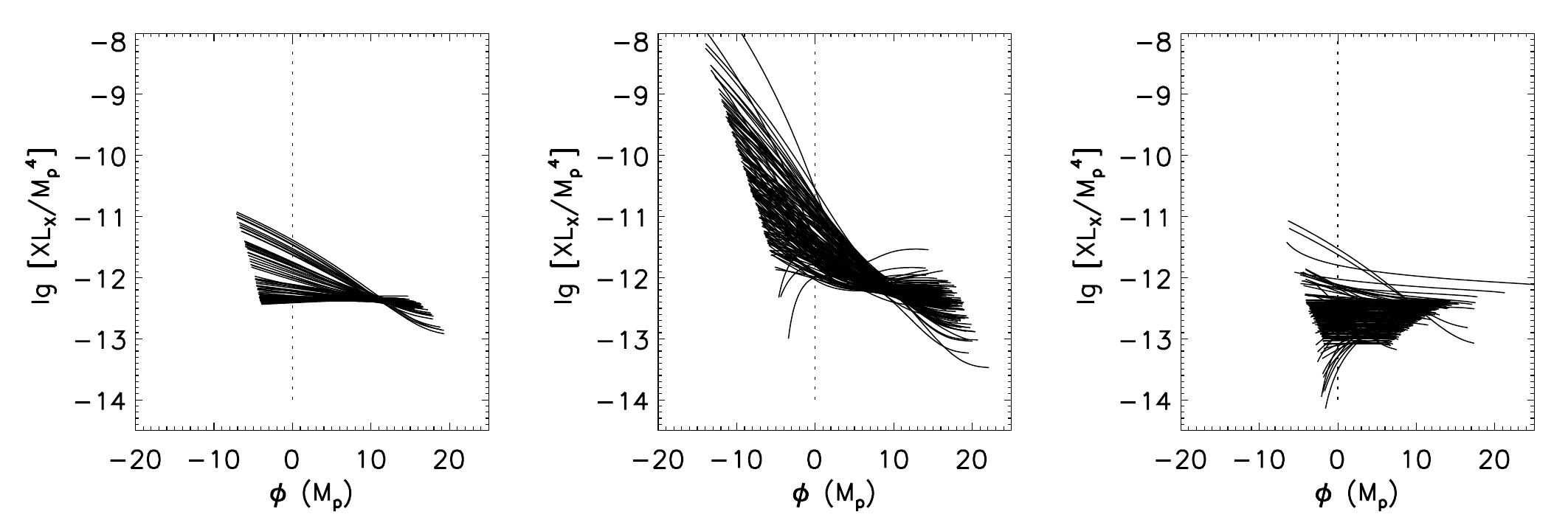}}
\caption{Reconstructed energy density $\rho$ [upper panels] and kinetic term $X{\cal L}_X$ [lower panels] for canonical models C1 [left panel], C2 [center] and general inflationary model G1 [right] which satisfy both conditions 1 and 2 from Sec. \ref{Four}. Each has $A_s, n_s$, $n_{run}$ at $k=k_*$ and the tensor-to-scalar ratio consistent with observations at the 2$\sigma$ level, and $\epsilon=1$ at the end of inflation with $N_*\sim 50-80$. The range of $\phi$ represents 100 e-foldings of evolution, with $\phi=0$ at scalar horizon crossing for $k_*$. (For canonical models with the usual scalar field definition, ${\cal L}_{X}=c_s^{-1}$, $X{\cal L}_{X}  =X$.)}
\label{fig10}
}
\end{figure*} 
%==========================================================================

%%%%%%%%%%%%%%%%%%%%%%%%%%%%%%%%%%%%%%%%%%%%%%
\subsection{Reconstruction  of viable inflationary trajectories}
\label{Fourc}
%%%%%%%%%%%%%%%%%%%%%%%%%%%%%%%%%%%%%%%%%%%%%%

In this section we consider the viable trajectories that satisfy both conditions 1 and 2, namely the spectral properties are consistent with observational constraints in the $k$ range measurable by CMB and large scale structure experiments, and inflation ends with $\epsilon=1$ around 60 e-foldings after observable modes have exited the horizon.

Fig. \ref{fig10} shows the results of $\sim 150$ reconstructed energy density evolutions $\rho(\phi)$, and kinetic term evolutions $X{\cal L}_{X}(\phi)$, tracing back 100 e-foldings from the end of inflation, for canonical and general models C1, C2, and G1, in which the power spectrum properties, $A_s, n_s, n_{run}$ at $k_*$, and the tensor-to-scalar ratio, are consistent with observations at the 2$\sigma$ level and $\epsilon=1$ at the end of inflation. We set $\phi=0$ at scalar horizon crossing for $k_*$, and $\phi$ at other epochs is given by
\bea
	\phi(N) & = & - \int_{N_*}^{N} \sqrt{2c_s(N')\epsilon(N')} dN'.
\eea

We see that introducing higher order flow parameters and/or the possibility of an evolving sound speed open up the range of allowed trajectories. Equally, however, observational constraints within the single field inflationary formalism are already starting to tie down the range of allowed inflationary histories for both canonical and general inflation.

%%%%%%%%%%%%%%%%%%%%%%%%%%%%%%%%%%%%%%%%%%%%%%
\section{Conclusions}
\label{Five}
%%%%%%%%%%%%%%%%%%%%%%%%%%%%%%%%%%%%%%%%%%%%%%

The large amount of data available today from CMB and large-scale structure surveys can be used to turn around the problem of matching inflationary theory to observations, to reconstructing the theory from the data itself. In this paper we explore the parameter space of general single field inflationary models by using a Monte-Carlo-Markov-Chain approach in combination with the Hubble flow formalism, and constrain the energy density and kinetic energy in such a Lagrangian. 

In order to accommodate models that allow the speed of sound, $c_{s}$, to vary during inflation, it is important to take into account the fact that tensor and scalar modes cross the horizon at different epochs. This directly affects the tensor-to-scalar ratio, which may deviate by as much as 50\% from the value obtained by assuming that horizon crossing epochs are effectively simultaneous.

We use the full flow parameter evolution equations to solve for the scalar and tensor perturbations spectra, and subsequently evolve each spectrum until it freezes out, as opposed to evaluating it at horizon crossing, in order to get precise predictions for the primordial power spectra over the full range of observable scales. We study five different classes of models of inflation, summarized in Table \ref{Table1}, in light of the latest CMB and large-scale structure data and show the observational constraints on the flow parameters, observed power spectrum, and typically considered observables, $n_s$, $n_{run}$, and $r$, at the pivot point $k=0.01$ Mpc$^{-1}$. 

Including higher-order slow-roll parameters allows a higher dimensional fit to the primordial power spectrum, with increased power on large and small scales possible in comparison to the commonly considered canonical power spectrum with scale independent running of the spectral index. In models of general inflation, where we allow $c_{s}$ to vary, the condition that inflation should continue (i.e. $\epsilon < 1$) on observable scales imposes a natural bound on the value of $\epsilon$ and $\kappa$ at horizon crossing of the pivot mode, and constrains the value of $r$.

In the absence of a sound theoretical explanation for inflation, the method of action reconstruction holds a lot of promise to give valuable directions for the search of such a theory. We impose constraints on the energy density and kinetic energy terms in the Lagrangian in light of current observations. The next step could now be to either explicitly construct a general class of allowed Lagrangians, or study different possible Lagrangians in light of these constraints. Once we have a general Lagrangian that could consistently explain the observed inflationary properties, the elucidation of its theoretical motivation should hopefully be that much closer.

%%%%%%%%%%%%%%%%%%%%%%%%%%%%%%%%%%%%%%%%%%%%%%
\section*{Acknowledgements}
%%%%%%%%%%%%%%%%%%%%%%%%%%%%%%%%%%%%%%%%%%%%%%

The authors would like to thank Daniel J.H. Chung, Ghazal Geshnizjani, William H. Kinney, Istvan Laszlo, Larissa C. Lorenz, Jerome Martin, Brian A. Powell, Sarah Shandera, S.-H. Henry Tye, and Alexander Vikman for helpful discussions during the course of this work. The authors compared results for standard inflation with that from the code generously made publicly available by the authors of \cite{Lesgourgues:2007aa}. N.A.'s and R.B.'s work is supported by NASA ATP Grant No. NNX08AH27G, NSF Grant Nos. AST-0607018 and PHY-0555216 and Research Corporation.

\bibliographystyle{apsrev}
\bibliography{inflation}

\end{document}